\documentclass{aa}
\usepackage{graphicx}
\usepackage{txfonts}
\usepackage{natbib}
\bibpunct{(}{)}{;}{a}{}{,}
\usepackage{hyperref}
\hypersetup{
    colorlinks=true,
    linkcolor=blue,
    citecolor=blue,
    urlcolor=blue
}

\usepackage[dvipsnames]{xcolor}

\usepackage{float}
\usepackage{booktabs}
\usepackage{caption}
\begin{document}

\title{Comparison between axisymmetric numerical magnetohydrodynamical simulations and self-similar solutions of jet-emitting disks}

\author{N. Zimniak\inst{1}, C. Zanni\inst{2} \and J. Ferreira\inst{1}}

\institute{Univ. Grenoble Alpes, CNRS, IPAG, 38000 Grenoble, France\\
    \email{nathan.zimniak@univ-grenoble-alpes.fr}
    \and
    INAF - Osservatorio Astrofisico di Torino, Strada Osservatorio 20, Pino Torinese 10025, Italy
}

\date{Received 28 October 2025 / Accepted 7 February 2026}

\abstract
{Turbulent accretion disks threaded by a large-scale vertical field near equipartition can drive tenuous and fast self-confined jets. Self-similar solutions of these jet-emitting disks (JEDs) have been known for a long time and provide the distributions of all physical quantities, from the turbulent disk to the asymptotic regime of ideal magnetohydrodynamic (MHD) jets. However, a thorough comparison with time-dependent numerical simulations has never been achieved, mostly because mass-loss rates found in simulations were always larger than those found analytically. This tension may have cast doubt on the analytical approach, the numerical one, or both.}
{Our goal is to bridge the gap between these two complementary approaches and settle this long-standing issue.}
{We performed 2.5D (axisymmetric) simulations of resistive and viscous accretion disks described by the same parameter sets as analytical JED solutions. The turbulent transport coefficients, as well as a turbulent magnetic pressure, have been included in the simulations. They follow the same $\alpha$ prescriptions as in the analytical work and are consistent with our current understanding of magnetic turbulence in accretion disks.}
{The numerical and the analytical solutions agree almost perfectly, and the previous tension now resolved. The JED solution thus appears to be structurally stable. Self-similarity is shown to bias the jet collimation properties only beyond the fast-magnetosonic point. Up to that point, the same set of disk parameters give rise to nearly indistinguishable numerical and analytical solutions, with magnetic surfaces displaying a near parabolic shape.The simulations also confirm that JEDs behave as dynamical attractors: starting from different initial conditions, the system consistently converges toward the expected steady-state solution.}
{This work demonstrates that self-similar solutions provide valuable insights into accretion-ejection physics. However, as 2.5D numerical simulations which rely on $\alpha$ prescriptions, they strongly depend on the assumptions made for turbulent terms. In contrast, 3D simulations capture the turbulence, but become prohibitively expensive when modeling large-scale astrophysical systems. We advocate for the use of global 3D simulations to investigate turbulence and to derive physically motivated prescriptions for use in 2.5D studies.}


\titlerunning{Comparison between MHD simulations and analytical JED solutions}
\authorrunning{N. Zimniak et al.}

\maketitle

\section{Introduction}

Accretion disks are ubiquitous in astrophysical systems, found around low-mass young stellar objects (YSOs, \citealt{Manara2023}); around supermassive black holes (BHs) in active galactic nuclei (AGNs) and quasars \citep{Liu2022}; in interacting binary systems containing a white dwarf (cataclysmic variables), a neutron star, or a BH (X-ray binaries) \citep{Mukai2017, Done2007}; and even around post-asymptotic giant branch (post-AGB) stars \citep{VanWinckel2018}. Highly collimated bipolar supersonic jets are observed to emerge from these systems. Detected mainly in the radio domain for compact objects \citep{Hardcastle2020} and from radio to optical emission lines in YSOs and binary post-AGB stars \citep{Ray2021}, these outflows consistently originate in the innermost regions, close to the central object.

Observations point to an interdependence between the two processes: mass accretion onto a central object despite near-Keplerian rotational equilibrium, and the liberation of mass in the form of two oppositely directed supersonic jets \citep{Coriat2011, Ghisellini2014, Nisini2018}. It is arguably the seminal work of \citet{Blandford1982}, which successfully explained jet acceleration and collimation, that defined the current accretion–ejection paradigm. Within this framework, a large-scale vertical magnetic field is assumed to thread the accretion disk over a significant region, providing a torque that leads to accretion and drives bipolar jets.

Self-consistent steady-state solutions of this accretion–ejection process were first obtained semi-analytically almost 30 years ago \citep{Ferreira1995,Ferreira1997,Casse2000a}. Using a self-similar ansatz, making it possible to keep all dynamical terms of the magnetohydrodynamic (MHD) equations, complete solutions have been successfully obtained. Such solutions make it possible to reconstruct 2.5D (axisymmetric) distributions of all physical quantities in both the disk and the jets. They also provide the parameter space for stationarity, expressed as the dependency $\xi(\mu)$, where $\xi$ is the local disk ejection efficiency (Eq.~\ref{eq:xi}) and $\mu$ the disk magnetization (Eq.~\ref{eq:mu}). However, this dependency is subject to several a priori specified conditions: the disk aspect ratio $\varepsilon \!=\! h/r$, where $h(r)$ is the local hydrostatic scale height based on the total kinetic (gas and radiation) pressure, which is used as a proxy for the true scale height of the disk; and several dimensionless parameters describing the MHD turbulence prevailing in the disk.

In these early self-similar models, only a near-equipartition magnetic field could steadily lead to supersonic\footnote{Strictly speaking, super-fast-magnetosonic jets, in the conventional sense (see, e.g., \citealt{Ferreira2004} and references therein).} jets. The disks associated with these highly magnetized solutions $(\mu \in [0.1, 1.0])$ have been labeled jet-emitting disks (JEDs; \citealt{Ferreira2006a}); accretion is supersonic, with tenuous and fast jets carrying away most of the released accretion power. Later on, weakly magnetized solutions $(\mu \ll 1)$ were also found \citep{JacqueminIde2019}, in which the magneto-rotational instability (MRI; \citealt{Balbus1991,Balbus2003}) plays an important role. In these weakly magnetized disks, accretion is highly subsonic, the vertical magnetic pinching is much less stringent, and more mass is actually ejected, leading to dense and slower collimated outflows. While these solutions have been labeled wind-emitting disks (WEDs) for convenience, there is a continuity in terms of disk magnetization, $\mu$, between JEDs and WEDs, in the same way that there is a continuity between the so-called magneto-centrifugally-driven jets \citep{Blandford1982} and magnetic towers \citep{LyndenBell2003}.

Despite their mathematical consistency, self-similar solutions have not been fully embraced by the community for two main reasons. The first concerns their existence and stability. Even though Newtonian gravity is self-similar, the question arises as to why astrophysical systems would adopt a self-similar configuration, and if they did, if it would be stable. The second reason relates to the prescriptions used for the disk MHD turbulence. Steady-state solutions require the disk plasma to diffuse across the vertical magnetic field, a process made possible only by turbulence, whereas jets can be treated using ideal MHD. Thus, although turbulent viscosity –as in a standard accretion disk (SAD; \citealt{Shakura1973})– is no longer required in a JED, turbulent magnetic diffusivity remains absolutely essential \citep{Ferreira1993a,Ferreira1993b}. However, we do not know how confident we can be in the turbulence prescriptions used in self-similar solutions. Their validity can be investigated through comparisons with time-dependent numerical experiments.

Since its rediscovery, the nonlinear stage of the MRI, namely MRI-driven turbulence, has been extensively studied using increasingly powerful (and costly) numerical tools. For almost 20 years, the primary goal was to characterize the anomalous viscosity coefficient $\alpha_v$, mostly through local (shearing-box) simulations (see, e.g., \citealt{Pessah2007,Salvesen2016}). It was eventually confirmed that $\alpha_v$ reaches its highest values in the presence of a vertical magentic field, $B_z$, with nonzero net magnetic flux, and it became clear that such configurations also inevitably lead to winds \citep{Suzuki2009,Fromang2013,Bai2013}. Since the MRI wavelength scales as $\lambda_{MRI} \! \propto \! \sqrt{\mu} h$, as the disk magnetization, $\mu$, increases the wind mass-loss rate increases, thereby requiring the use of global 3D simulations to fully capture the dynamics \citep{Lesur2013,Suzuki2014,JacqueminIde2019}.

Performing global 3D ideal MHD simulations that capture both small-scale (turbulent) and large-scale (accretion–ejection) spatio-temporal dynamics requires a substantial effort, one that has been accomplished only very recently \citep{Zhu2018,JacqueminIde2021,JacqueminIde2024,Tu2025}. Surprisingly, weakly magnetized disks threaded by a large-scale vertical magnetic field, $B_z$, appear to develop an elevated structure, reaching heights on the order of $r$, even though their thermal scale remains $h \!=\! \varepsilon r \ll r$ (see also \citealt{Lancova2019}). This vertical expansion is caused by strong turbulent magnetic pressure, mostly related to the toroidal magnetic field $B_\phi$ (see \citealt{JacqueminIde2021} and references therein). MRI-driven turbulence not only generates turbulent magnetic pressure in addition to anomalous viscosity, it also provides an anomalous magnetic diffusivity with a complex vertical profile within the disk \citep{Zhu2018,JacqueminIde2021}. Taken together, these results sound the death knell for the SAD model and raise significant concerns for semi-analytical WED models, as current self-similar solutions assume turbulent coefficients that decrease monotonically with height.

In contrast, 3D numerical simulations of strongly magnetized disks appear to exhibit turbulent profiles that decrease with height. However, this remains difficult to confirm from the existing literature (see, e.g., \citealt{Scepi2024a}), for several reasons. First, strong magnetic fields are computationally much more challenging to handle than weak fields, and only global simulations performed in the framework of general relativity (GRMHD) have been reported so far, with the notable exception of \citealt{Mishra2020}. In GRMHD studies, the focus has primarily been on the plunging region around Kerr (rotating) black holes and on the formation of electromagnetic jets along the axis \citep{Blandford1977}. As a result, no study has yet analyzed disk turbulence in the Newtonian regime (beyond ten gravitational radii), let alone compared numerical results with analytical models, even though mass loss in the form of winds is systematically observed (see, e.g., \citealt{McKinney2006,Tchekhovskoy2010,Tchekhovskoy2011,McKinney2012,Avara2016,Narayan2022}). In this context, strongly magnetized disks are referred to as magnetically arrested disks (MADs), a term originally based on incorrect early expectations \citep{Narayan2003,Igumenshchev2003}. It is now understood that MADs are likely the numerical counterpart of semi-analytical JEDs, the disk vertical balance preventing (before the radial one) the magnetic field from reaching an overwhelmingly large value \citep{Ferreira2022,Scepi2024b,Zimniak2024}.

Comparing JED solutions with the results of 3D numerical simulations of strongly magnetized disks first requires measuring the final disk magnetization, $\mu$, which is not straightforward; most studies report only the plasma beta, $\beta$, which is calculated using all the magnetic-field components, both laminar and turbulent. If turbulence were negligible, one would simply have $\mu\!=\!2/\beta$ at the disk midplane, but this is unfortunately not the case \citep{Mishra2020,Scepi2024a}. However, by combining the available information from these studies and making some reasonable assumptions, it is possible to estimate $\mu \!\sim\! 0.1$ in certain simulations, placing them at the lower end of the cold JED parameter space (see \citealt{Zimniak2024} and Sect.~5). In these strong-field simulations, the turbulence appears to decrease more steeply with height than in weak-field cases, the disk no longer being puffy. Still, without a dedicated analysis of turbulence and jet properties in strongly magnetized disks from 3D simulations, it remains difficult to pursue a more detailed comparison.

An alternative approach consists of solving the same set of equations and turbulence prescriptions using a time-dependent 2.5D (axisymmetric) MHD code. This is much easier and less computationally expensive, although it obviously cannot provide any insight into MHD turbulence itself. However, it offers the advantage of enabling the computation of accretion–ejection structures on much larger spatio-temporal scales, comparable to those observed. It also makes it possible to test the validity of the self-similar mathematical approximation, identify potential biases, and, more importantly, assess whether the solutions are structurally stable. Such 2.5D simulations have been performed for both strongly magnetized disks (JEDs), with $\mu$ being between 0.2 and 1.0 \citep{Casse2002,Zanni2007,Tzeferacos2009,Sheikhnezami2012}, and weakly magnetized disks (WEDs), with $\mu$ being as low as $10^{-4}$ \citep{Murphy2010,Sheikhnezami2012,Stepanovs2014,Stepanovs2016}. These simulations confirmed the general accretion–ejection dynamics, namely the interplay of forces that allows accretion to proceed simultaneously with the generation of cold\footnote{Cold jets, in contrast to warm jets \citep{Casse2004,Tzeferacos2013}, occur when enthalpy plays no role in jet launching. Heat deposition at the disk surface can significantly enhance the ejection efficiency, $\xi$ \citep{Casse2000b}.} jets, exactly as predicted by the analytical solutions. However, the disk mass-loss rate was consistently found to be much higher than that predicted by self-similar solutions. For this reason, no precise comparison has ever been carried out between the two.

The goal of this work is to perform 2.5D numerical MHD simulations and compare the results with semi-analytical self-similar solutions, not only in terms of parameter space, but also by examining the vertical profiles of all physical quantities in greater detail. Since MHD turbulence is known to generate a turbulent magnetic pressure, we included this additional term in our $\alpha$-disk prescription. We then compared the numerical outcomes to the new self-similar solutions of \citet{Zimniak2024}, which incorporate this effect.

Section~2 presents the physical framework used to describe a magnetized accretion–ejection structure, along with the numerical method employed to solve it. Section~3 describes the results of our numerical simulations and analyzes the impact of turbulent magnetic pressure. Section~4 then provides a detailed comparison between these new solutions and the self-similar JED solutions, showing an almost perfect agreement between the two for the first time. The origin of previous discrepancies with earlier simulations is discussed in Section~5. Finally, our conclusions are summarized in Section~6.

\section{Numerical accretion-ejection model}

\subsection{MHD equations and the numerical method}

We modeled the time-dependent interplay between a non-relativistic plasma, organized as a turbulent accretion disk orbiting a central object of mass $M$ and a large-scale vertical magnetic field that threads the disk. We modeled the disk turbulence by including an anomalous viscosity, magnetic diffusivity, and a turbulent magnetic pressure term. The set of MHD equations is as follows:

\vspace{-\medskipamount}

\begin{equation}
\begin{aligned}
\label{eq_MHD}
\frac{\partial \rho}{\partial t} & + \nabla\cdot\left(\rho \vec {v}\right) = 0 \\
\frac{\partial \rho\vec{v}}{\partial t} & + \nabla \cdot \left[ 
\rho \vec{v}\vec{v} +
\left( P + P_{turb} + \frac{\vec{B}\cdot\vec{B}}{2\mu_0} \right)\vec{I}-
\frac{\vec{B}\vec{B}}{\mu_0} - \vec{\bar{ \bar{ \mathcal{T}} }}
\right]  = - \rho \nabla \Phi_G \\
\frac{\partial {\cal E}}{\partial t} & + \nabla\cdot\left[
\left({\cal E} + P + \frac{\vec{B}\cdot\vec{B}}{2\mu_0}\right)\vec{v}-
\frac{\left(\vec{v}\cdot\vec{B}\right)\vec{B}}{\mu_0} \right] =  \\
& = \vec{v} \cdot \left( \nabla \cdot \vec{\bar{ \bar{ \mathcal{T}} }} \right)
- \frac{\vec{B}}{\mu_0} \cdot \left(\nabla \times \vec{\bar{\bar{\eta}}_m} \vec{J}\right) - \vec{v} \cdot \nabla P_{turb} - \Lambda \\
\frac{\partial \vec{B}}{\partial t} & + \nabla \times \left(\vec{B}\times\vec{v} + \vec{\bar{\bar{\eta}}_m} \vec{J} \right)= 0 \; . 
\end{aligned}
\end{equation}

\noindent this accounts for the conservation of mass, momentum, and energy, as well as the induction equation. In these equations, $\rho$ is the plasma density, $\vec{v}$ the velocity, $P$ the thermal pressure, $P_{turb}$ the turbulent magnetic pressure, $\vec{B}$ the magnetic field, $\vec{J} \!=\! \nabla\times\vec{B}/\mu_0$ the electric current, $\mu_0$ the vacuum magnetic permeability, $\Phi_G \!=\! -GM/R$ the gravitational potential of the central object,  $\vec{\bar{\bar{\mathcal{T}}}}$ the viscous stress tensor, and $\vec{\bar{\bar{\eta}}_m}$ the turbulent magnetic resistivity tensor. The viscous stress tensor only includes the $\mathcal{T}_{R\phi} \!=\! \rho \nu_v \mathcal{D}_{R\phi}$, where $\vec{\bar{\bar{\mathcal{D}}}}$ is the usual displacement tensor. 
The total laminar (or bulk) energy density is defined as ${\cal E} \!=\! \rho u(T) + \rho \vec{v} \!\cdot\! \vec{v}/2 + \vec{B} \!\cdot\! \vec{B}/(2 \mu_0) + \rho \Phi_G$ and is the sum of the laminar internal, kinetic, magnetic and gravitational potential energy densities. 
In the associated energy conservation equation, the turbulent contributions, incorporated through $\alpha$-type parametrizations discussed in the next section, can affect the bulk kinetic energy (through the work exerted by the anomalous viscous stress and the turbulent magnetic pressure) and the bulk magnetic energy (through the turbulent diffusion of magnetic energy), but they cannot modify the internal energy (e.g. Ohmic and viscous heating terms are not included).
Note that, for simplicity, neither radiation losses nor any turbulent heat transport term (see, e.g., \citealt{Shakura1978}, sect.~4.2 in \citealt{Ferreira1995}, and \citealt{Casse2000b}) were included since we focused on (cylindrically) isothermal disks and outflows. In order to achieve this condition, a heating–cooling function, $\Lambda$, was included, the definition of which is given in the next section. 
Due to our simplified treatment of the thermodynamics, we did not introduce any phenomenological ad hoc prescription for the exchange between turbulent pressure (or energy) and internal energy. These choices are consistent with our assumption of a locally isothermal flow (see Section 2.2), for which no separate evolution equation for the turbulent energy is required. 

The definition of the specific internal energy $u(T)$ as a function of temperature, $T$, is provided in Appendix A of \citet{Pantolmos2020}. This equation of state allowed us to model a calorically imperfect gas, meaning that the polytropic index $\gamma = C_P/C_V$, i.e., the ratio between the specific heats at constant pressure and volume, is not constant, but it can depend on temperature. The plasma behaves almost isothermally at high temperatures and adiabatically at low temperatures, giving the possibility to model at the same time a hot axial wind/spine and a cold disk wind. We set the equation of state to have $\gamma = 1.05$ for $P/\rho > 0.1 \,. GM/R_0$ and $\gamma = 5/3$ for $P/\rho < 0.025 \, GM/R_0$, where $R_0$ is the inner radius of the disk and the computational domain. 

Beside the system of Eqs.~(\ref{eq_MHD}), we also solved a passive scalar equation:
\begin{equation}
\frac{\partial \rho T_{\!r}}{\partial t} + \nabla \cdot \left( \rho T_{\!r}\, \vec{v} \right) = 0 \; ,
\label{eq_trac}
\end{equation}
where $T_{\!r}$ is a passive tracer that was used to track and separate the disk and disk-wind material (with $T_{\!r} = 1$) from the axial spine material (with $T_{\!r} = 0 $). The value of the tracer in each cell actually provides the density fraction of material coming from the disk.

We solved the system of equations (\ref{eq_MHD}) using the PLUTO code \citep{Mignone2007} and a finite volume approach. Time integration was carried out with a second-order Runge-Kutta (RK2) scheme. To compute intercell fluxes, we used the HLLD Riemann solver \citep{Miyoshi2005} together with the van Leer slope limiter \citep{vanLeer1979}. The divergence-free condition $\vec{\nabla} \cdot \vec{B} \!=\! 0$ was preserved at machine precision using the constrained transport method \citep{Evans1988}. 
The simulations were performed in 2.5D, i.e., all the components of the 3D vector fields are computed by assuming an axial symmetry. We solved the equations in a spherical system of coordinates, $[R, \theta]$, and we used the notation $[r,z] = [R\sin\!\theta, R\cos\!\theta]$ for the cylindrical coordinates. The computational domain is defined over the radial interval $R \in [R_0, 535 R_0]$, where $R_0$ is the radius of the inner spherical boundary, and over the polar interval $\theta \in [0, \pi/2]$. The domain was discretized using a grid with $N_R \!=\! 512$ points in the radial direction and $N_{\theta} \!=\! 156$ points in the polar direction. The radial grid is logarithmically spaced, providing increased resolution in the inner regions of the disk, where the steady-state regime is established first. The polar grid is mostly uniform, but it becomes progressively refined near the equatorial plane to ensure high resolution within the disk.

\subsection{Numerical setup}
\label{sect:Ns}

In the initial conditions, we set up a slightly sub-Keplerian disk threaded by a purely poloidal magnetic field, where the thermal and turbulent disk pressure and the centrifugal acceleration balance the equilibrium against the radial gravitational pull and the vertical magnetic field pinch. Above the disk we initially set a hot thermal atmosphere that, with suitable boundary conditions, can support an almost isothermal Parker-like wind emerging from the inner boundary to fill the axial region inside the disk-wind. In order to distinguish the two outflow components, we imposed the value of the passive tracer Eq. (\ref{eq_trac}) to $T_{\!r} = 1$ to mark the disk and the disk-wind material 
and tell it apart from the coronal or stellar wind launched from the inner boundary identified by a value of $T_{\!r} = 0$. All the details about initial and boundary conditions are provided in Appendix~\ref{sect:A}. 

The disk's initial condition is characterized by a constant thermal aspect ratio, $\epsilon = h/r$, is defined by the ratio between the midplane isothermal sound speed; i.e., $c_s = \sqrt{P / \rho} = \sqrt{2kT / m_p}$, the Keplerian speed $V_K = \sqrt{GM/r}$, and a constant magnetization of

\begin{equation}
    \mu = \frac{V_A^2}{c_s^2} = \frac{B^2}{\mu_0 P} ,
    \label{eq:mu}
\end{equation}

defined by the square of the ratio between the Alfv\'en $V_A = B_z/\sqrt{\mu_0 \rho}$ and the isothermal sound speed $c_S$ on the disk equatorial plane. The heating–cooling function in the energy equation of the MHD system (\ref{eq_MHD}) was set to

\begin{equation}
    \Lambda = \rho \, C_V \, T_{\!r} \, \frac{T-T_{\textrm{eff}}}{\tau} \; ,
    \label{eq:Lambda}
\end{equation}

where $C_V$ is the specific heat capacity at constant volume defined in Appendix A in \citet{Pantolmos2020}, $T_{\textrm{eff}}$ is a relaxation temperature and $\tau$ is the corresponding relaxation timescale. We set $T_{\textrm{eff}}$ to the initial midplane disk temperature, $T \propto 1/r$, and the timescale $\tau = 0.1 /\sqrt{GM/r^3}$ to a fraction of the local Keplerian period, so that the accretion–ejection solution relaxes to a vertically isothermal structure, with the temperature set by the initial disk temperature. Notice that the function Eq. (\ref{eq:Lambda}) is also multiplied by the tracer $T_{\!r}$ so that it did not affect the quasi-isothermal structure of the hot thermal spine that evolves according to the assumed equation of state. 

We assumed an $\alpha$ prescription to define the turbulent terms using the same form as in the analytical solutions (see \citealt{Zimniak2024} for more details). The turbulent magnetic diffusivity tensor $\vec{\bar{\bar{\eta}}_m}$ is assumed to be diagonal, with $\eta_{m_{RR}} \!=\! \eta_{m_{\theta\theta}} \!=\! \mu_0 \nu_m$ and $\eta_{m_{\phi\phi}} \!=\! \mu_0 \nu_m / \chi_m$, where $\chi_m$ is the turbulence anisotropy parameter. In this study, we restricted ourselves to the isotropic case $\chi_m = 1$ for simplicity. The turbulent magnetic diffusivity at the disk midplane follows the prescription of \citet{Ferreira1993a}:
\begin{equation}
\nu_m = \alpha_m V_A h \; ,
\end{equation}
where $\alpha_m$ is an adimensional parameter, $V_A \!=\! B_z / \sqrt{\mu_0 \rho}$ is the midplane Alfvén speed and $h = c_s/\Omega_K$ is the thermal height scale of the disk. Notice that while the value of the equatorial Alfvén speed can change with time, the thermal height scale, $h$, stays constant due to the presence of the function $\Lambda$ from Eq. (\ref{eq:Lambda}). The turbulent viscosity, $\nu_v \!=\! \mathcal{P}_m \nu_m$, is related to the turbulent magnetic diffusivity, $\nu_m$, via the (effective) magnetic Prandtl number, $\mathcal{P}_m$. The definition of the disk's equatorial magnetization $\mu \!=\! (V_A / c_s)^2$ leads to the well-known scaling for the viscosity coefficient $\alpha_v \!=\! \alpha_m {\cal P}_m \mu^{1/2}$ found in MRI studies, with typical values of $\alpha_m \mathcal{P}_m$ ranging from a few to around eight (\citealt{Salvesen2016,JacqueminIde2021} and references therein). Notice that the disk magnetization is not necessarily constant with time, since the midplane Alfvén speed can change, while the sound speed stays approximately constant. For simplicity, and to facilitate comparison with analytical studies, we adopted $\mathcal{P}_m \!=\! \alpha_m \!=\! 1$ here. 

Turbulent magnetic pressure generated by MRI can be highly anisotropic, but, due to the ordering of $\lvert\, \delta B_\phi \rvert \gg \lvert\, \delta B_r \rvert \gg \lvert\, \delta B_z \rvert$, we only considered fluctuations of the toroidal magnetic field and employed the scaling found in MRI studies \citep{Salvesen2016}: 

\begin{equation}
    P_{turb} \simeq \frac{ \langle \delta B_\phi^2 \rangle}{2\mu_0} = \alpha_P \sqrt{\mu} P \ , 
\end{equation}
where $P$ is the thermal pressure which provides an isotropic turbulent pressure as a first-order approximation.

As in \citet{Zimniak2024}, we explored several values of the coefficient $\alpha_P$. To remain consistent with self-similar analytical studies, and to confine turbulence to the disk only, all these transport coefficients –as well as the turbulent magnetic pressure– were multiplied by a Gaussian profile: $f(x) \!=\! \exp[-(x/x_t)^2/2]$, where $x \!=\! z/h$, or equivalently $f(\theta) = \exp[-(\cos\theta/\sin\theta \epsilon x_t)^2/2]$, where $x_t$ is the vertical scale height of turbulence stratification. In all simulations presented here, we set $x_t \!=\! 1/\sqrt{2}$. As a result, the ideal MHD regime (with negligible turbulent magnetic pressure) is approximately recovered one thermal scale height above the disk midplane.

In summary, the initial conditions depend on nine dimensionless quantities. We fixed seven of them: the disk thermal height scale, $\epsilon = 0.1$; the turbulent magnetic diffusivity coefficient, $\alpha_m = 1$; its anisotropy, $\chi_m = 1$; the effective Prandtl number, $\mathcal{P}_m=1$; the turbulent height scale, $x_t = 1/\sqrt{2}$; the injection sound speed of the axial wind, $c_{s0}/\sqrt{GM/R_0} = 0.39$; the initial density contrast between the midplane disk density and the axial spine density at $R=R_0$, $\rho_{d0}/\rho_{s0} = 1000$. Our simulations are therefore defined by different values of the initial disk magnetization, $\mu$, and the turbulent pressure amplitude, $\alpha_P$.

\subsection{Units and normalization}

As is customary, simulations were performed and results are presented in dimensionless units. Lengths are given in units of the radius of the inner boundary $R_0$, corresponding to the internal radius of the disk. Given the mass of the central object, $M$, velocities are presented in units of the Keplerian speed, $V_{K0} = \sqrt{GM/R_0}$, while time is given in units of the inner Keplerian period, $t_0 = 2\pi/\sqrt{GM/R_0^3}$. Expressing the density in units of the density of the axial wind at the inner boundary, $\rho_{s0}$ –which we remind the reader is a factor of 1000 smaller than the initial disk density at the same radius– the mass accretion and ejection rates are given in units of $\dot{M}_{0} = \rho_{s0} V_{K0} R_0^2 $. For example, assuming values typical for a solar-mass, young forming star with $\rho_{s0} = 10^{-15} \, \mathrm{g\ cm}^{-3}$, $M = M_\sun$, and $R_0 = 0.1\ \mathrm{AU}$ we obtain
\begin{equation}
\begin{aligned}
   V_{K0} &= 94 \left(\frac{M}{M_{\sun}}
   \right)^{1/2} \left(\frac{R_0}{0.1\ \mathrm{AU}}  \right)^{-1/2}\ \
   \mathrm{km\ s^{-1}}\\
   t_{0} &= 11.5 \left(\frac{M}{M_{\sun}}\right)^{-1/2}
   \left(\frac{R_0}{0.1\ \mathrm{AU}}  \right)^{3/2}\ \ \mathrm{days} \\
   \dot{M}_{0} &= 3.3 \times 10^{-10}
   \left(\frac{\rho_{s0}}{\scriptstyle 10^{-15}\ \mathrm{g}\
   	\mathrm{cm}^{-3}}  \right)
   \left(\frac{M}{M_{\sun}}\right)^{1/2}
   \left(\frac{R_0}{0.1\ \mathrm{AU}}  \right)^{3/2}\ \
   M_{\sun}\ \mathrm{yr^{-1}} \; .
\end{aligned}
\end{equation}

The simulations were carried out for 1000 Keplerian orbits at the inner disk radius. This timescale ensures that the system is evolved through a sufficient number of inner orbits to allow the innermost regions to reach a steady-state regime. Time is discretized into $N_t \!=\! 1000$ uniformly spaced steps across the interval, enabling accurate resolution of both transient dynamics and long-term behavior. 

\begin{figure*}
  \centering
  \resizebox{12cm}{!}{\includegraphics{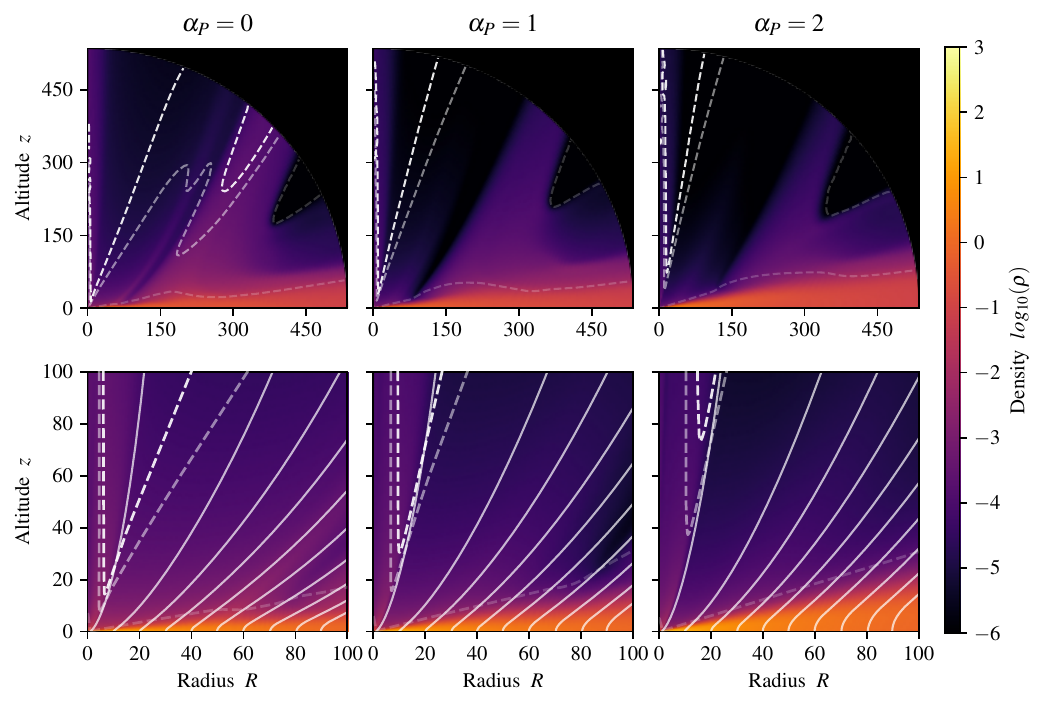}}
  \hfill
  \parbox[b]{55mm}{
    \caption{Two-dimensional visualizations at $t = 1000$ for simulations with $\alpha_P = 0$, $1$, and $2$, from left to right. The top panels display the full computational domain, while the bottom panels show a zoomed-in view of the inner region. Dashed lines, with increasing opacity, represent the slow magnetosonic (SM), Alfvén, and fast magnetosonic (FM) critical surfaces, respectively. White lines indicate magnetic-field lines anchored at $R = 2, 10, 20, 30$, and up to $90$. The background color map shows the density in arbitrary units.}
    \label{fig:fig1}
  }
\end{figure*}

\section{Numerical simulations}

\begin{table*}
     \centering
     \caption{Initial parameters $(\alpha_P, \mu_i)$ and final quantities at $t \!=\! 1000$ for each simulation.}
     \begin{tabular}{lcc||cccccccccccc}
        \toprule
        Name & $\alpha_P$ & $\mu_{i}$ & $R_{ss}$ & $\mu$ & $\xi$ &$m_s$  & $\delta$ & $\kappa$ & $\lambda$ & $\omega_*$ & $e$ & $\dot{M}_{acc}$ & $\dot{M}_{jet}$ & $\dot{M}_{spine}$ \\
        \midrule
        \midrule
        \#1 & 0.0 & 0.6 & 13.2 & 0.7 & 0.0087 & 2.8 & 0.8 & 0.0084 & 55  & 0.73 & 78  & 122.2 & 1.32 & 2.14 \\
        \#2 & 1.0 & 2.0 & 4.5  & 2.0 & 0.0028 & 2.8 & 0.7 & 0.0012 & 187 & 0.65 & 243 & 164.8 & 0.29 & 1.40 \\
        \#3 & 2.0 & 4.0 & 2.3  & 4.0 & 0.0010 & 2.3 & 0.6 & 0.0004 & 359 & 0.63 & 449 & 202.0 & 0.09 & 0.78 \\
        \bottomrule
        \vspace{-2mm}
     \end{tabular}
    \tablefoot{Quantities listed are: the outer radius $R_{ss}$ of the self-similar domain, disk magnetization $\mu$, ejection index $\xi$, sonic Mach number $m_s$, deviation $\delta$ from Keplerian rotation (all evaluated at $R \!=\! 10$), normalized MHD invariants $\kappa$, $\lambda$, $\omega_*$, $e$ (measured along a field line anchored at $R \!=\! 10$), accretion rate $\dot{M}_{acc}$ (measured at $R_{ss}$), disk ejection rate $\dot{M}_{jet}$ (measured between $R_{in}$ and $R_{ss}$), and spine ejection rate $\dot{M}_{spine}$. All simulations are performed with $\varepsilon \!=\! 0.1$, $\alpha_m \!=\! \chi_m \!=\! \mathcal{P}_m \!=\! 1$, and $x_t \!=\! 1 / \sqrt{2}$.}
    \label{tab:tab1}
\end{table*}

We present the results of three simulations, each characterized by a unique combination $(\alpha_P, \mu_i)$ of the turbulent magnetic pressure intensity, $\alpha_P \in [0, 1, 2]$, and the initial disk magnetization, $\mu_i \in [0.6, 2.0, 4.0]$, respectively. The list of parameters is provided in Table~\ref{tab:tab1}.

The parameter sets $(\alpha_P, \mu_i)$ were chosen based on the theoretical solutions of \citet{Zimniak2024} (hereafter Z24), in order to favor faster convergence, if any, toward these solutions. This approach allows us not only to assess whether the numerical simulations recover the same final magnetization $\mu$, but also to evaluate the influence of the turbulent magnetic pressure parameter $\alpha_P$. In the following, we provide a detailed analysis of the overall structure and dynamical evolution of the simulations, focusing on the magnetic field configuration, the establishment of a steady-state regime, and the properties of both the disk and the jet. In parallel, we examined how the turbulent magnetic pressure affects the system as $\alpha_P$ increases.

\subsection{Global picture}

Figure~\ref{fig:fig1} shows snapshots at $t \!=\! 1000$ for the three simulations, with the full computational domain displayed in the top panels and a zoomed-in view shown in the bottom panels. All simulations appear to have reached a steady state up to a certain cylindrical radius. Accretion is primarily driven by the laminar torque exerted by the jet. The ejected material successively becomes super-slow magnetosonic (super-SM, first dashed line close to the disk), then super-Alfvénic (second dashed line), and finally super-fast magnetosonic (super-FM) for streamlines anchored at sufficiently small radii.

The fact that the inner region reaches a steady state in which all three critical surfaces (SM, A, and FM) are almost conical is a strong indication that self-similarity is preserved, despite the presence of an axial spine (visible as the zone where the A and FM surfaces rise along the axis). This result is consistent with previous simulations of the same kind. It is also worth noting that once the FM surface reaches the outer boundary of the domain, continuing the simulation becomes unnecessary: extending the steady-state region to a larger radial extent in the disk would require enlarging the spherical domain. As will be shown below, the final stages of the simulations closely match the self-similar solutions.

As the intensity of turbulent magnetic pressure, $\alpha_P$, increases, the magnetic-field lines become increasingly vertical, and the critical surfaces are found at progressively smaller colatitude angles; i.e., the flow reaches SM, A, and FM velocities at higher altitudes. This behavior is consistent with the trends observed in the analytical studies of Z24 and can be interpreted as a consequence of the enhanced poloidal magnetic-field strength associated with increasing $\alpha_P$, which tends to push the critical surfaces farther from the disk. An increase in $\alpha_P$ also leads to higher values of the accretion rate $\dot{M}_{acc} \!=\! 4 \pi R^2 \int_0^{\theta_d} \rho v_R sin(\theta) d \theta$ (with $\theta_d$ being the location where the radial [spherical] velocity vanishes), and lower values of the disk ejection rate, $\dot{M}_{jet} \!=\! -2 \pi \int_{R_0}^{R_{ss}} R \rho v_\theta sin(\theta) dR$ (with $R_{ss}$ being the last radius where the flow is in a stationary state; see below for more details), and the spine ejection rate, $\dot{M}_{spine} \!=\! 4 \pi R^2 \int_{\theta_s}^{\pi/2} \rho v_R sin(\theta) d \theta$ (where $\theta_s$ denotes the polar angle of the spine surface, defined as the location where $T_r=0.1$). For all three simulations, mass conservation is satisfied following $2 \dot{M}_{jet} = \dot{M}_{acc}(R_{ss}) - \dot{M}_{acc}(R_{in})$, with an error of less than $20\%$.

\subsection{Steady-state regime}

\begin{figure}
    \centering
    \includegraphics[width=0.9\columnwidth]{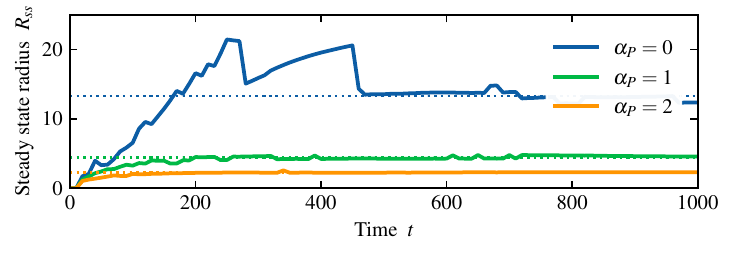}
    \caption{Time evolution of the anchoring radius $R_{ss}$ of the last magnetic field line intersecting the FM surface for each simulation. The dashed curves represent the final stationary radii: $R_{ss} \!=\! 13.3$ for $\alpha_P \!=\! 0$, $R_{ss} \!=\! 4.5$ for $\alpha_P \!=\! 1$ and $R_{ss} \!=\! 2.3$ for $\alpha_P \!=\! 2$.}
    \label{fig:fig2}
\end{figure}

To compare the simulations with steady-state theory, a precise criterion for identifying the steady regime must first be defined. One option is to examine how the disk accretion rate, $\dot M_{acc}(r)$, evolves over time and to consider only the innermost region where it remains stationary. Alternatively, since accretion is ultimately driven by angular momentum removal by jets, it may be more appropriate to verify whether the MHD invariants remain constant along magnetic-field lines. Both criteria are computed and verified in Appendix~\ref{sect:B}.

As noted by \citet{Ferreira1997}, the dynamical timescale for the flow to become super-FM (in the conventional sense) is on the order of the Keplerian period at the anchoring radius, $r_o$, of the magnetic surface. However, it takes significantly longer for a transverse equilibrium to settle in and establish the final shape of the FM critical surface. This process requires a redistribution of mass and magnetic flux within the disk, and it is therefore strongly dependent on the initial conditions. Our choice of initial configuration was designed to minimize transients as much as possible, leading to the establishment of a self-similar domain, $\mathcal{D}_{ss}$, where the critical surfaces are conical. The extent of this domain is quantified by the anchoring radius, $R_{ss}$, of the outermost magnetic surface for which the flow becomes super-FM.

Figure~\ref{fig:fig2} shows the time evolution of $R_{ss}$ for the three simulations: $R_{ss} \!=\! 13.3$ is reached at $t_{ss} \!=\! 450$ for $\alpha_P \!=\! 0$, $R_{ss} \!=\! 4.5$ at $t_{ss} \!=\! 200$ for $\alpha_P \!=\! 1$, and $R_{ss} \!=\! 2.3$ at $t_{ss} \!=\! 100$ for $\alpha_P \!=\! 2$. The decrease in $R_{ss}$ with increasing $\alpha_P$ is already qualitatively observed in Figure~\ref{fig:fig1} and reflects the fact that FM surfaces become more vertical as turbulent magnetic pressure increases. It is, however, interesting to see that the time required to reach transverse equilibrium does not strictly follow a Keplerian scaling. Although the $\alpha_P \!=\! 0$ simulation converges more rapidly (i.e., with a larger slope), it still takes longer to reach steady state due to the larger radial extent involved.

\subsection{Radial structure}

\begin{figure*}
  \centering
  \resizebox{12cm}{!}{\includegraphics{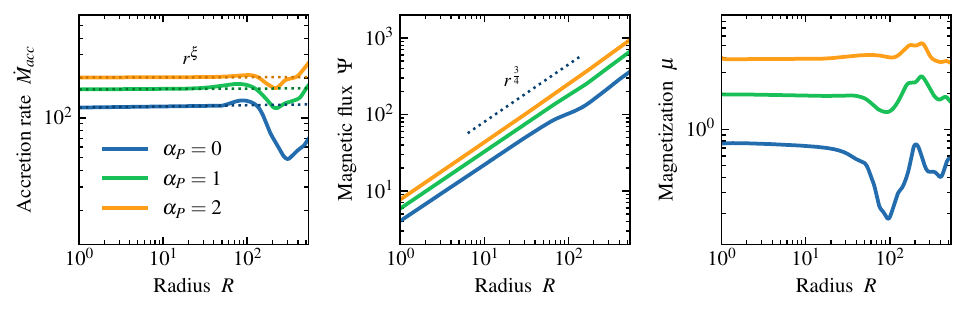}}
  \hfill
  \parbox[b]{55mm}{
    \caption{Radial distributions of disk accretion rate, $\dot{M}_{acc}$; magnetic flux, $\Psi$; and magnetization, $\mu$; measured at the disk midplane ($\theta \!=\! \pi/2$) for each simulation at $t \!=\! 1000$. Dashed lines indicate simple power-law fits as functions of the cylindrical radius, $r$ (see text for details).}
    \label{fig:fig3}
  }
\end{figure*}

Figure~\ref{fig:fig3} displays the radial distributions of several disk quantities for each simulation at $t \!=\! 1000$. These quantities follow power-law profiles consistent with the analytical solutions beyond the previously derived transverse steady-state radius, $R_{ss}$ (note that $t \!=\! 1000$ corresponds to one orbital period at $r \!=\! 100$).

This confirms that the jet transverse structure takes significantly longer to adjust than the disk (see also Appendix~\ref{sect:B}). In steady-state analytical models, the disk magnetization, $\mu$, is constant, the angular velocity scales as $\Omega \propto r^{-3/2}$, and the other quantities follow:

\vspace{-\medskipamount}

\begin{eqnarray}
    \dot M_{acc}(r) \propto r^\xi \ , & \rho \propto r^{\xi - 3/2} \ , & B_z \propto r^{-5/4 + \xi/2}
    \label{eq:xi}
\end{eqnarray}

\noindent leading to a magnetic flux of $\Psi \!=\! \int_{1}^r B_z 2 \pi r dr \propto r^{3/4 + \xi/2}$. We fit the accretion rate over the interval $R \in [1,10]$ for all three simulations and obtain $\xi \!=\! 0.0087$ for $\alpha_P \!=\! 0$, $\xi \!=\! 0.0028$ for $\alpha_P \!=\! 1$, and $\xi \!=\! 0.0010$ for $\alpha_P \!=\! 2$. These extremely small values indicate that the initial radial dependencies imposed on $B_z$ (and therefore on $\Psi$) and on $\rho$ are effectively preserved. Alternative methods to estimate the ejection index, $\xi$, are discussed below, but we emphasize that the present values are consistent with the analytical models.

Moreover, a JED with nearly constant magnetization is established in the innermost region of each simulation. The values, measured at $R \!=\! 2$, are $\mu \!=\! 0.7$ for $\alpha_P \!=\! 0$, $\mu \!=\! 2.0$ for $\alpha_P \!=\! 1$, and $\mu \!=\! 4.0$ for $\alpha_P \!=\! 2$. Again, these values closely match the corresponding initial conditions, indicating that the numerical JEDs have converged toward solutions very similar to those of Z24.

\subsection{Latitudinal structure}

\begin{figure*}
  \centering
  \resizebox{12cm}{!}{\includegraphics{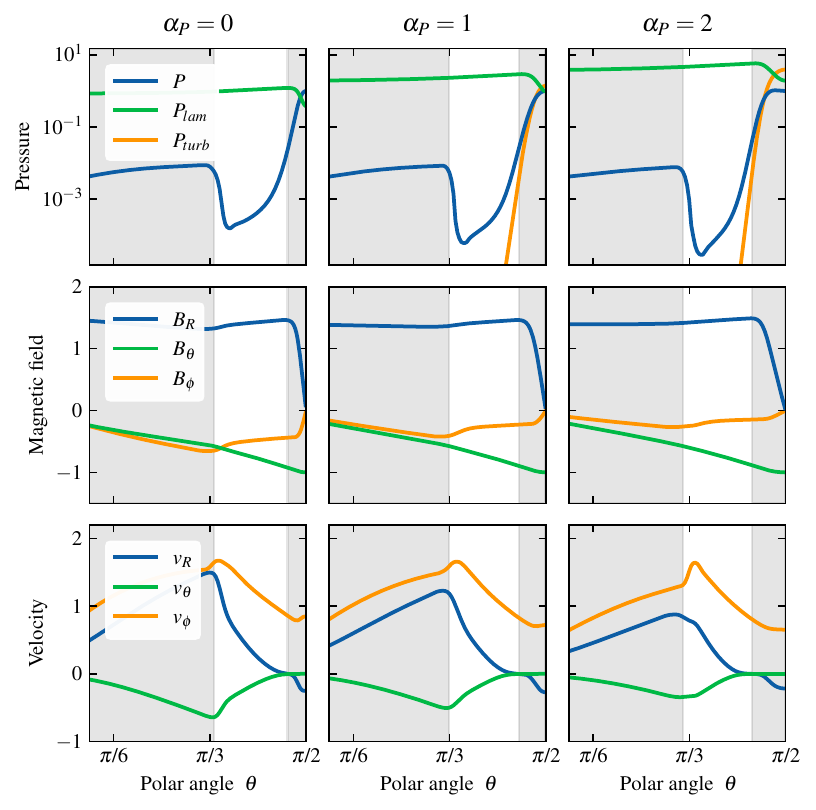}}
  \hfill
  \parbox[b]{55mm}{
    \caption{Latitudinal profiles of thermal pressure $P$; laminar (total) magnetic pressure, $P_{lam}$; turbulent magnetic pressure, $P_{turb}$ (all normalized to $P$ at the disk midplane); components of the magnetic field $B_R$, $B_{\theta}$, and $B_{\phi}$ (normalized to $B_z$ at the disk midplane), and components of the velocity $v_R$, $v_{\theta}$, and $v_{\phi}$ (normalized to the Keplerian velocity $v_{K}$ at the disk midplane), measured at $R \!=\! 2.0$ and at $t \!=\! 1000$ for each simulation. The gray shaded area on the right corresponds to the disk, the white area to the disk wind, and the gray shaded area on the left to the spine.}
    \label{fig:fig4}
  }
\end{figure*}

Figure~\ref{fig:fig4} shows the latitudinal profiles of various disk quantities for the three simulations, evaluated at $R \!=\! 2$.

The pressures (kinetic $P$, laminar magnetic $P_{lam} \!=\! B^2/2\mu_0$ and turbulent $P_{turb}$) were normalized to $P$ at the disk midplane. The laminar magnetic-field components were normalized to $B \!=\! B_z \!=\! ||B_\theta||$ at the midplane. The velocity components were normalized to the Keplerian velocity at the midplane. These latitudinal profiles are easy to extract from simulations but slightly cumbersome to understand as one goes from the spine (gray zone below $\theta \!\simeq\! \pi/3$), crossing several field lines in ideal MHD (white zone in the middle of the plots), until reaching an angle of $\theta_d$ defining the resistive accretion disk down to the midplane at $\theta \!=\! \pi/2$ (gray zone). In our analysis, we define the $\theta_d$ angle as the location where the radial (spherical) velocity vanishes, which separates the ejecting zone above from the accreting region below. The axial gray zone that defines the spine was identified using a passive tracer included in PLUTO. This allowed us to determine, at each time step, the origin of material injected at the inner radial boundary.

As expected for JEDs, the dominant torque is due to the jets and leads to a supersonic accretion speed at the disk midplane. Defining the sonic Mach number as $m_s \!=\! -v_R / c_s$ at the midplane, with the sound speed given by $c_s \!=\! \Omega_K h$, the simulations provide $m_s \!=\! 2.8, 2.8,$ and $2.3$ for $\alpha_P \!=\! 0,1,$ and $2$, respectively, at $R \!=\! 2$ (but this is valid throughout the self-similar domain). As the magnetization, $\mu$, increases, the disk becomes progressively more sub-Keplerian. Using the $\delta \!=\! \Omega / \Omega_K$ ratio at the midplane, we find $\delta \!=\! 0.8$, $0.7$, and $0.6$ for the three simulations. It is worth noting that the rotation rate continues to decrease toward the disk surface due to the negative torque exerted by the jet; then, it starts to increase above it, eventually reaching super-Keplerian values once the torque becomes positive. This behavior is associated with magnetic acceleration, as described in \citet{Ferreira1995}.

\begin{figure*}
  \centering
  \resizebox{12cm}{!}{\includegraphics{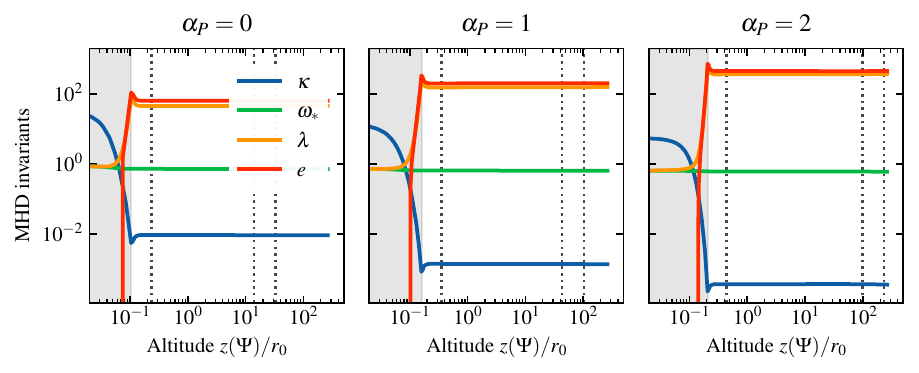}}
  \hfill
  \parbox[b]{55mm}{
    \caption{Profiles of MHD invariants along magnetic-field line anchored at $R \!=\! 2.0$ for each simulation at $t \!=\! 1000$. The quantities shown are the mass-loading parameter, $\kappa$; the normalized angular velocity of the magnetic surface, $\omega_*$; the magnetic lever arm, $\lambda$; and the normalized total specific energy, $e$.
    The shaded area indicates the resistive MHD region, while the dashed vertical lines mark the positions of the SM, Alfvénic, and FM critical surfaces.}
    \label{fig:fig5}
  }
\end{figure*}

In summary, despite the presence of the axial spine, the numerical simulations reproduce the same trends with $\alpha_P$ as those predicted by the analytical solutions presented in Z24.
In particular, as the turbulent magnetic pressure increases, the disk becomes puffier (the gray zone increases), and the toroidal magnetic field, $B_\phi$, at the disk surface becomes smaller. This reduction in $B_\phi$ weakens the magnetic torque due to the jets, thereby reducing their power\footnote{Note, however, that as the jet becomes more collimated, the axial spine also becomes more compressed, since both outflows must remain in transverse pressure balance.}. A decrease in the outward radial-velocity component, $v_r$, can already be observed at this stage. To assess jet properties, one must, however, look instead of the MHD invariants.

\subsection{MHD invariants}

A steady-state, axisymmetric outflow governed by ideal MHD is structured with magnetic surfaces nested around each other. Each surface carries at least four conserved quantities, known as MHD invariants \citep{Blandford1982, Pelletier1992}: the mass-to-flux ratio, $\eta(\Psi) \!=\! \mu_0 \rho v_p / B_p$, the specific total angular momentum, $L(\Psi) \!=\! \Omega r^2 - r B_\phi / \eta$; the angular velocity of the magnetic surface, $\Omega_*(\Psi) \!=\! \Omega - \eta B_\phi / (\mu_0 \rho r)$; the total specific energy (Bernoulli invariant), $E(\Psi) \!=\! u^2/2 + H + \Phi_G - \Omega_*rB_\phi/\eta + E_{turb}$, where $H \!=\! (\gamma/\gamma-1)P/\rho$ is the enthalpy; and $E_{turb}$, the contribution of turbulent magnetic pressure (which is negligible in our case; see Z24). A fifth invariant may exist depending on the jet thermodynamics. However, this is not relevant here since the jets are cold and the enthalpy term $H$ is also negligible.

For practical purposes, it is convenient to define normalized versions of these invariants: the mass-loading parameter, $\kappa \!=\! \eta B_o / (\Omega_{K_o} r_o)$; the magnetic lever arm, $\lambda \!=\! L / (\Omega_{K_o} r_o^2)$; the surface rotation, $\omega_* \!=\! \Omega_* / \Omega_{K_o}$; and the Bernoulli invariant, $e \!=\! 2E / (\Omega_{K_o}^2 r_o^2)$. Here, the subscript "o" refers to quantities evaluated at the disk midplane and anchoring radius $r_o$. With these definitions, the maximum asymptotic poloidal speed along a given magnetic surface is given by $u_{p,\infty} \!=\! \Omega_{K_o} r_o \sqrt{e}$. For a cold jet, accretion–ejection theory provides the relation $\lambda \!\simeq\! \omega_* + q / \kappa \!\simeq\! \omega_* + 1 / (2\xi)$, where $q \!\simeq\! -B_\phi / B_z$ is the magnetic shear at the disk surface. The specific energy then satisfies $e \!\simeq\! 2 \omega_* \lambda - 4 + \omega_*^2$ (see Z24). These expressions highlight the central role of the disk ejection efficiency $\xi$ in jet dynamics.

Figure~\ref{fig:fig5} shows the profiles of the four normalized MHD invariants along the magnetic-field line anchored at $r_o \!=\! 2.0$ for each simulation at $t \!=\! 1000$. The three vertical dashed lines, ordered from the disk outward, indicate the locations of the SM, Alfvén, and FM critical surfaces. The gray shaded region corresponds to the resistive MHD disk, where the invariants are not defined. In the ideal MHD regime, the four quantities $\kappa$, $\lambda$, $\omega_*$, and $e$ remain constant along the magnetic-field line, as expected. Their values are reported in Table~\ref{tab:tab1}. The analytical approximation $\kappa \!\sim\! \xi$, valid for cold jets in the absence of turbulent magnetic pressure \citep{Ferreira1997}, breaks down as $\alpha_P$ increases. In contrast, the relation $e(\lambda, \omega_*)$ remains accurate, with deviations below 2\%. The simplified expression $\lambda \!\simeq\! 1 + 1 / (2\xi)$ continues to yield estimates within 5 to 10\% for JEDs.

The global structure obtained in numerical simulations is in excellent agreement with the corresponding analytical solutions; the effect of increasing the turbulent disk magnetic pressure, $\alpha_P$, on jet dynamics follows the analytical findings of Z24 exactly. Higher magnetization, $\mu$, leads to a decrease in the ejection efficiency, $\xi$, which in turn produces larger magnetic lever arms, $\lambda$, even though the rotation rate of the magnetic surfaces, $\omega_*$, becomes slower. The next step is to extend this comparison in a more quantitative manner.

\section{Comparison with self-similar models}

\subsection{Methodology}

\begin{figure*}
  \centering
  \resizebox{12cm}{!}{\includegraphics{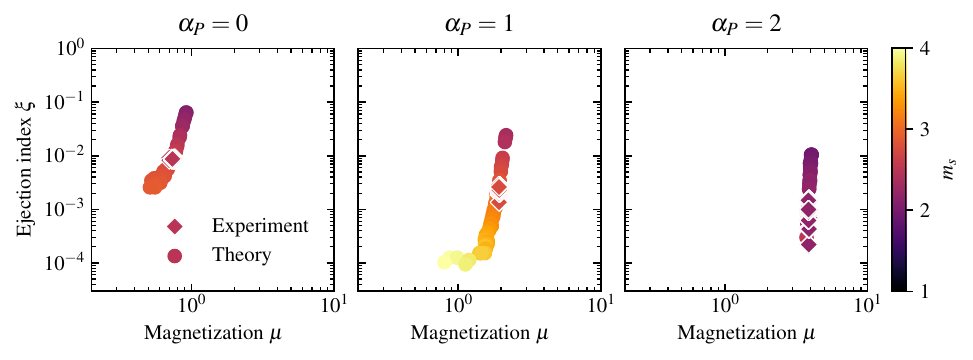}}
  \hfill
  \parbox[b]{55mm}{
    \caption{Comparison between disk parameter spaces $(\xi, \mu)$ obtained from numerical simulations (diamonds) and from theoretical self-similar solutions (dots) for different values of the turbulent magnetic pressure parameter $\alpha_P$.
    Both the simulations and the self-similar models produce super-FM jets and are based on the same set of turbulence parameters (see text).}
    \label{fig:fig6}
  }
\end{figure*}

\begin{figure*}
  \centering
  \resizebox{12cm}{!}{\includegraphics{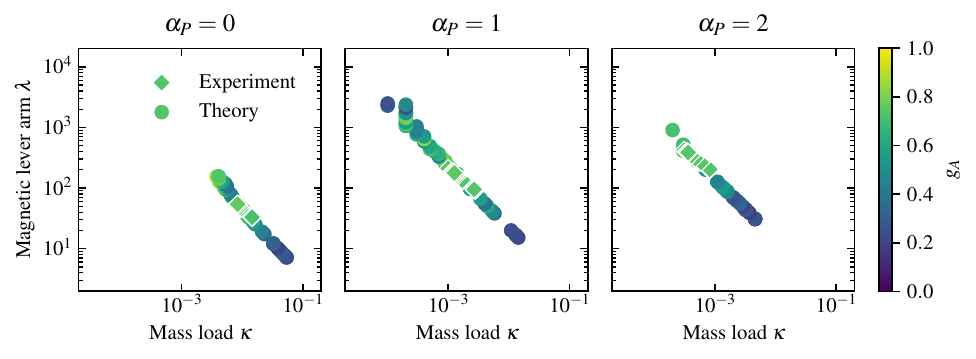}}
  \hfill
  \parbox[b]{55mm}{
    \caption{Comparison between disk parameter spaces $(\lambda, \kappa)$ obtained from numerical simulations (diamonds) and from theoretical self-similar solutions (dots) for different values of the turbulent magnetic pressure parameter $\alpha_P$.
    Both the simulations and the self-similar models produce super-FM jets and are based on the same set of turbulence parameters (see text).}
    \label{fig:fig7}
  }
\end{figure*}

Our objective was to identify, for each simulation, a self-similar analytical solution that reproduces the same set of disk parameters as closely as possible. This required us to first select a pair $(\xi, \mu)$ with associated turbulence parameters and then compare the resulting vertical profiles (disk and jet). To achieve this, we used the JED parameter space explored in Z24 for different values of $\alpha_P$, as shown in Fig.~\ref{fig:fig6}. This figure reveals a certain degree of degeneracy, since a rather wide range of $\xi$ is possible for a tiny (almost none for $\alpha_P \!=\! 2$) variation in $\mu$. This indicates that the local disk ejection efficiency, $\xi$, is not uniquely determined by the local magnetization, $\mu$, it also depends on the radial distributions of quantities such as $B_z$ and $\rho$, and therefore on the boundary conditions imposed.

For this purpose, we built a much larger set of numerical stationary outcomes than the three analyzed previously. Indeed, at every radius within the self-similar domain $R_{ss}$ and for all times beyond $t_{ss}$ (as defined in Sect.~3.2), all disk and jet quantities can be evaluated. The time interval between two simulation outputs was $\Delta t \!=\! 50$, which is long enough to ensure local stationarity across the domain below $R_{ss}$. This enabled us to construct a collection of local equilibria, which defined our "experimental" data set $(\xi, \mu)$ for each value of $\alpha_P$. Given its importance, the ejection efficiency, $\xi$, was computed as the average of two independent estimates. The first was obtained by fitting a radial power law of $r^\xi$ to the disk accretion rate, $\dot{M}_{acc}(r)$, over the interval $[R_{in}, R_{ss}]$. The second was derived from steady-state mass conservation $2\dot{M}_{jet}/\dot{M}_{acc}(R_{ss}) \!=\! 1- (R_{ss}/R_{in})^\xi$, where $\dot{M}_{jet}$ is the total mass lost from the jet. Figure~\ref{fig:fig6} displays the juxtaposition of our experimental (diamonds) and theoretical (dots) parameter spaces in the $(\xi, \mu)$ plane for different values of the turbulent magnetic pressure intensity $\alpha_P$. The colors indicate the sonic Mach number, $m_s$, at the disk midplane. The agreement is excellent for all values of $\alpha_P$.

The same methodology was applied to the ideal MHD jet quantities. Figure~\ref{fig:fig7} presents the juxtaposition of our experimental (diamonds) and theoretical (dots) parameter spaces in the $(\lambda, \kappa)$ plane for different values of the turbulent magnetic pressure intensity $\alpha_P$. Once again, the two sets show excellent agreement, including in the color scale, which represents the quantity of $g_A$ measured at the Alfvén surface. This quantity, defined by $\Omega \!=\! \Omega_*(1 - g)$, describes how the plasma rotation, $\Omega$, deviates from the rotation rate of the magnetic surface $\Omega_*$ \citep{Pelletier1992}. At the Alfvén surface, $g_A \!\simeq\! r_A B_{\phi_A} / (r_o B_\phi)$ quantifies the fraction of the poloidal electric current still available for further acceleration of the jet. The value of $g_A$ influences the altitude of the Alfvén surface, $z_A$, and plays a key role in determining the degree of jet collimation \citep{Ferreira1997, JacqueminIde2019}. All numerical solutions obtained here show $g_A \!\sim\! 0.6-0.7$, which suggests that these jets are able to propagate over large distances.

Based on the previous analysis, three self-similar analytical solutions from Z24 were selected to best reproduce three numerical solutions. The corresponding parameters for each case are listed in Table~\ref{tab:tab2}.

\subsection{Vertical profiles}

\begin{table}
    \centering
    \caption{Main parameters of numerical simulations and the corresponding semi-analytical solutions used for comparison.}
    \begin{tabular}{lccccc}
        \toprule
        Solution type & $\alpha_P$ & $\mu$ & $\xi$ & $\kappa$ & $\lambda$ \\
        \midrule
        \midrule
                        & 0.0 & 0.77 & 0.00878 & 0.00944 & 45.9 \\
        Numerical       & 1.0 & 1.96 & 0.00289 & 0.00140 & 158.0 \\
                        & 2.0 & 3.86 & 0.00098 & 0.00037 & 372.2 \\
        \midrule
                        & 0.0 & 0.75 & 0.00876 & 0.00956 & 45.9 \\
        Semi-analytical & 1.0 & 1.92 & 0.00237 & 0.00163 & 139.6 \\
                        & 2.0 & 3.86 & 0.00081 & 0.00040 & 357.1 \\
        \bottomrule
        \vspace{-2mm}
    \end{tabular}
    \label{tab:tab2}
\end{table}

\begin{figure*}
  \centering
  \resizebox{12cm}{!}{\includegraphics{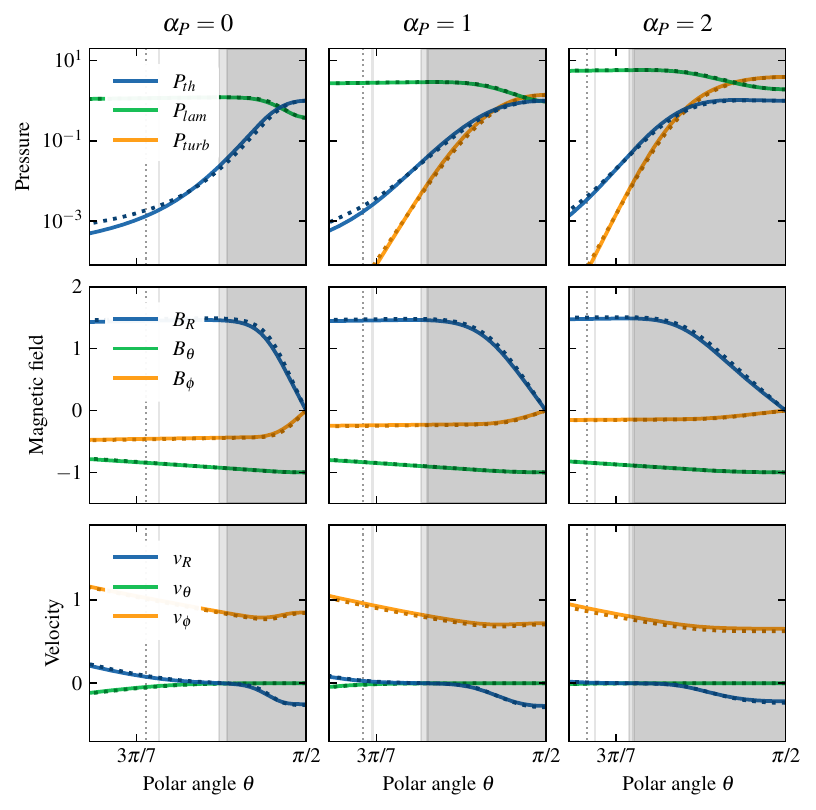}}
  \hfill
  \parbox[b]{55mm}{
    \caption{Latitudinal profiles (zoomed in on the disk) of thermal pressure, $P$; laminar (total) magnetic pressure, $P_{lam}$; turbulent magnetic pressure, $P_{turb}$ (all normalized to $P$ at the disk midplane); components of the magnetic field, $B_R$, $B_{\theta}$, and $B_{\phi}$ (normalized to $B_z$ at the disk midplane); and components of the velocity, $v_R$, $v_{\theta}$, and $v_{\phi}$ (normalized to the Keplerian velocity, $v_{K}$, at the disk midplane) from three numerical (solid lines) and theoretical (dashed) solutions. The shaded regions correspond to the disk (light gray for the numerical solution and dark gray for the semi-analytical one), and the vertical lines mark the transition to the ideal MHD regime (solid lines for the simulations and dashed lines for the semi-analytical solutions).}
    \label{fig:fig8}
  }
\end{figure*}

\begin{figure*}
  \centering
  \resizebox{12cm}{!}{\includegraphics{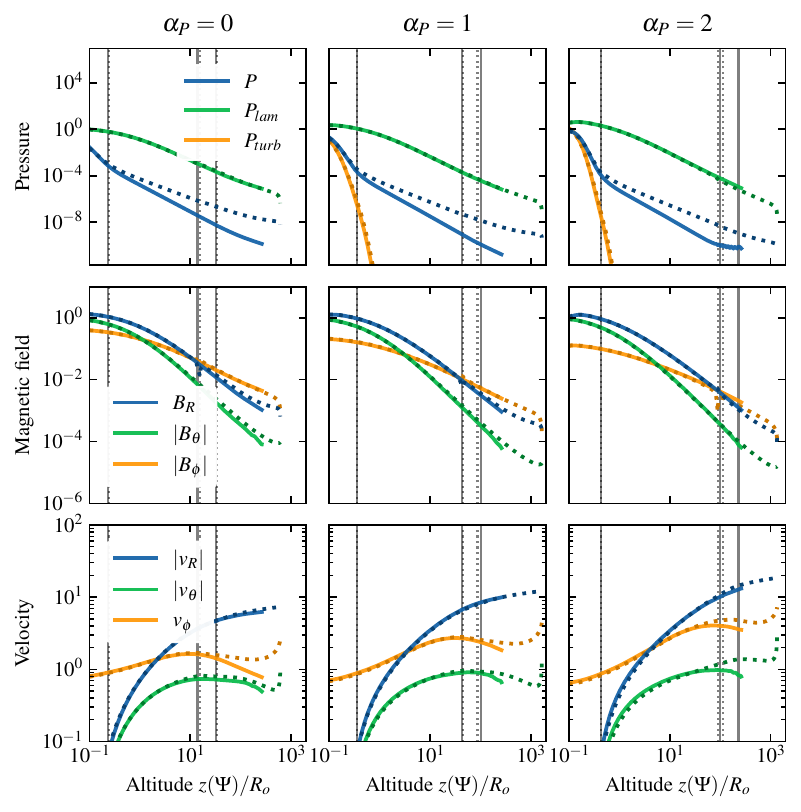}}
  \hfill
  \parbox[b]{55mm}{
    \caption{Same as in Fig.~\ref{fig:fig8}, but along a magnetic surface. The vertical lines indicate the locations of the critical points SM, Alfvén, and FM (solid lines for the simulations and dashed lines for the semi-analytical solutions).}
    \label{fig:fig9}
  }
\end{figure*}

Figure~\ref{fig:fig8} presents a zoomed-in view of the region near the resistive MHD disk, showing the latitudinal profiles obtained from the numerical simulations (solid lines) and the analytical self-similar solutions (dotted lines). The shaded areas indicate the extent of the accreting disk, with light gray representing the numerical solution and dark gray the analytical one. Vertical lines mark the transition to the ideal MHD regime, with solid lines for the simulation and dashed lines for the theory. In all cases, the SM critical point appears shortly after this transition. Both sets of profiles were normalized using the same conventions. The resulting near-perfect agreement strongly supports the conclusion that self-similar solutions accurately reproduce the numerical results. A slight difference can be observed in the thermal pressure, $P$, above the disk, which arises from differences in the thermodynamic treatment. Specifically, the self-similar solutions from Z24 assume isothermal conditions along magnetic surfaces, whereas the numerical simulations are isothermal along cylinders. As the magnetic surfaces open, the temperature and, thus, the pressure drop, explaining the divergence. However, since the jets are cold and the enthalpy contribution is negligible, this has no impact on the jet dynamics. Another minor difference lies in the altitude at which the transition to ideal MHD occurs. The semi-analytical approach benefits from significantly higher resolution, allowing the resistive MHD terms to extend farther from the disk midplane than in the simulations. This too had no significant effect on the resulting profiles. We do not comment any further on these curves as the underlying physics of the accretion–ejection process has already been thoroughly discussed in the theoretical literature.

Figure~\ref{fig:fig9} shows the profiles of the same physical quantities, this time along the magnetic-field line. Once again, the self-similar solutions reproduce the numerical solutions with remarkable accuracy, including the locations of the three critical points. There are, however, some discrepancies. First, the discrepancy in thermodynamic behavior increases as the jet widens, which is expected given the difference in assumptions and has no consequence for cold flows. Second, semi-analytical jets propagate significantly farther than those in the simulations, with a maximum altitude of $z_{max}/r_o \!=\! 614, 1665, 1351$ for the three analytical models –compared to $z_{max}/r_o \!=\! 263$ in the simulations– due to the finite size of the computational domain. Third, and most importantly, self-similar jets systematically recollimate toward the axis, as evidenced here by the increase in $v_\phi$, whereas the simulated jets continue to open outward. Among these, the last point represents the only significant deviation, and it warrants further investigation.

\begin{figure}
    \centering
    \includegraphics[width=\columnwidth]{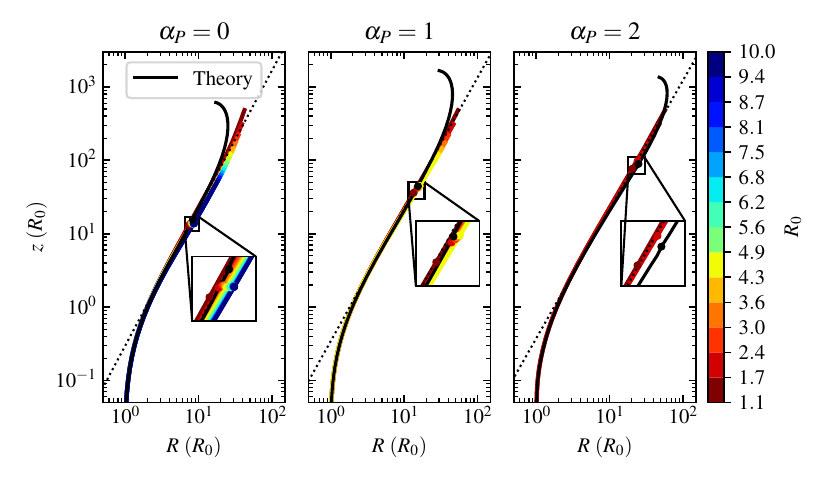}
    \caption{Shape of magnetic-field lines anchored between $r_o \!=\! 1.1$ and $r_o \!=\! 10$ (colored lines) obtained from the simulations, superimposed on the corresponding self-similar solutions (solid black lines). Insets show the location of the Alfvén points (dots) along each surface. The dotted line represents a power-law fit of the form $z/r_o = (r/r_o)^n - z_o/r_o$, where $r$ denotes the cylindrical radius (see text).}
    \label{fig:fig10}
\end{figure}

To better assess the differences in the asymptotic behavior of numerical and self-similar jets, Fig.~\ref{fig:fig10} shows the shapes of several magnetic-field lines anchored at radii between $r_o \!=\! 1.1$ and $r_o \!=\! 10.0$ (colored lines); these are superimposed on the corresponding self-similar solutions. Across most of the domain, the agreement between the numerical and theoretical magnetic-field line shapes is truly astonishing. It is noteworthy that all magnetic surfaces within this anchoring range display self-similar profiles. This holds even for $\alpha_P \!=\! 1, 2$, for which the stationary steady-state radii were estimated only up to $R_{ss} \!=\! 4.5, 2.3$, respectively. This provides an additional, albeit indirect, indication that the underlying disk has reached a stable configuration out to at least $r_o \!=\! 10$. The figure also reveals that, at least for $\alpha_P \!=\! 0$, the innermost magnetic-field lines would have had sufficient space to recollimate within the computational domain if they were inclined to do so. The absence of such recollimation thus emerges as the first and, to date, only clear divergence from self-similar predictions. Despite this difference, both the numerical and analytical jets follow a similar large-scale geometry. Over most of the domain, they can be accurately fit by a power law of the cylindrical radius $z/r_o = (r/r_o)^n - z_o/r_o$, with $n \!=\! 1.88, 1.80, 1.74$, and $z_o/r_o \!=\! 0.53, 0.47, 0.43$ as $\alpha_P$ increases. This confirms that MHD jets launched from strongly magnetized disks are approximately parabolic in shape.

Since recollimation in self-similar jets arises from the increasing dominance of the hoop stress as the jet opens up \citep{Ferreira1997}, its absence from the simulations implies a lower axial poloidal electric current. Indeed, in self-similar solutions, the toroidal magnetic field, $B_\phi$, diverges near the axis, whereas the axis in the simulations is occupied by a spine. This quenching effect caused by the axial spine has also been reported in platform simulations of jets, where recollimation still occurred but at much larger distances than predicted by analytical models \citep{Jannaud2023}.

\begin{figure*}
  \centering
  \resizebox{12cm}{!}{\includegraphics{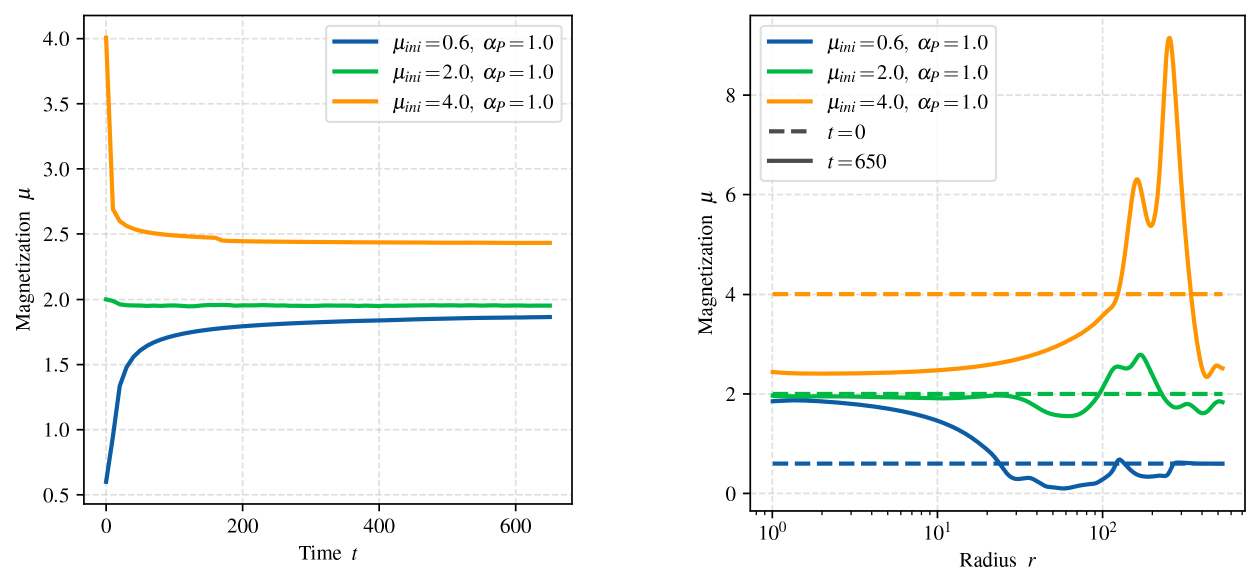}}
  \hfill
  \parbox[b]{55mm}{
    \caption{Left: Magnetization, $\mu$, measured at $R=2$ as a function of time, $t$, for three different initial magnetizations: $\mu_{i} \in [0.6, 2.0, 4.0]$ at $\alpha_P \!=\! 1.0$. Right: Radial profiles of magnetization, $\mu$, measured at $t=650$ for three different initial magnetizations: $\mu_{i} \in [0.6, 2.0, 4.0]$ at $\alpha_P \!=\! 1.0$.}
    \label{fig:fig11}
  }
\end{figure*}

\section{Discussion}

This work presents the first instance of an almost perfect agreement between the final stationary outputs of axisymmetric MHD simulations and steady-state self-similar accretion–ejection solutions. A key result, which had never been reproduced before, is the very small value of the disk ejection efficiency, $\xi \!\sim\! 0.01$, achieved for cold outflows from JEDs (near-equipartition magnetic field). Earlier studies reported significantly higher values: \citet{Casse2002} found $\xi \!\simeq\! 0.3$, \citet{Zanni2007} reported $\xi \!\simeq\! 0.125$, and \citet{Tzeferacos2009} obtained $\xi \!\simeq\! 0.036$. Although these results followed the expected trend toward lower values, $\xi$ consistently remained above the analytical prediction. This evolution can be attributed to both the progressive increase in resolution and the use of progressively less diffusive numerical schemes. Later simulations, even those performed at higher resolution, still yielded $\xi \!\sim\! 0.3$ because they employed a stronger magnetic diffusivity, both in amplitude, $\alpha_m$, and over a larger vertical extent, $m(x)$, which enhanced mass diffusion from the disk and prevented direct comparison with analytical studies \citep{Sheikhnezami2012, Stepanovs2014}.

We performed several numerical tests by degrading the resolution (from $16$ to two grid cells across the disk) and by modifying the numerical scheme (see Fig.~\ref{fig:figC1} in Appendix~\ref{sect:C}). These tests show that the correct value of the disk ejection efficiency $\xi$ can already be obtained at moderate resolution (with eight cells across the disk), which therefore cannot explain the larger values reported in, for example, \citet{Tzeferacos2009}. We suspect instead that the cylindrical grid employed in those earlier studies introduced a bias that has remained unnoticed. Indeed, Fig.~6 of \citet{Zanni2007}, Fig.~11 of \citet{Tzeferacos2009}, and Fig.~8 of \citet{Sheikhnezami2012}, reveal that the poloidal electric current enters the disk at its upper surface ($J_z < 0$, $J_r < 0$) rather than through its inner lateral edge. This feature results from the boundary conditions typically used with cylindrical coordinates. By construction, such a configuration produces an inner region where the radial electric current flowing outward from inside the disk increases with distance, giving rise to an outward vertical push of the mass by the Lorentz force that enhances mass ejection and leads to artificially large values of $\xi$ (see the discussion in \citealt{Ferreira1997} and the figures mentioned above). In contrast, our boundary conditions were specifically designed to ensure that the axial electric current enters the disk through its inner radial boundary (see Appendix~\ref{sect:A}).

An interesting aspect of JED configurations lies in the equatorial symmetry imposed in analytical solutions. Removing this symmetry is computationally demanding and, somewhat paradoxically, it was first done in 3D simulations, where some degree of top-down asymmetry was reported (see, e.g., \citealt{Bethune2017,JacqueminIde2021}); it was also recently seen in 2.5D simulations \citep{Tu2025}. Nevertheless, all these studies were done with initially very weak magnetic fields. In contrast, we performed a 2.5D simulation with $\alpha_P \!=\! 0$ over the full $\theta \in [0,\pi]$ domain and found no significant deviation from the corresponding equatorially symmetric case (see Appendix~\ref{sect:C}).

More importantly, our numerical study confirms the structural stability of JEDs. Indeed, we performed three simulations with different initial magnetizations, $\mu_{i} \in [0.6, 2.0, 4.0]$, while keeping the turbulent magnetic pressure fixed at $\alpha_P \!=\! 1$. These initial values correspond to those predicted by the semi-analytical solutions for different $\alpha_P$ values (see Fig.~\ref{fig:fig6}). Figure~\ref{fig:fig11} (left) shows the time evolution of the magnetization $\mu$, measured at $R=2$, for each simulation. In the $(\mu_{i} \!=\! 2.0, \alpha_P \!=\! 1)$ case, $\mu$ remains almost constant, as expected since the initial conditions are already very close to the semi-analytical solution. In the $(\mu_{i} \!=\! 0.6, \alpha_P \!=\! 1)$ case, the magnetic flux accumulates in the inner disk regions, and the structure stabilizes once the magnetization reaches $\mu \simeq 2.0$. In the $(\mu_{i} \!=\! 4.0, \alpha_P \!=\! 1)$ case, the magnetic flux is instead advected outward. The structure again appears to converge toward $\mu \!\simeq\! 2.0$, but the characteristic timescales become too large to reach this value without evolving the simulation over much longer times. Figure~\ref{fig:fig11} (right) shows the radial profiles of magnetization at the initial and final times for the same simulations. The results indicate that the magnetization converges over a radial range extending from $R_{in}$ to about $R \sim 10$. Beyond this radius, the simulations have not yet had sufficient time to dissipate transients and converge toward the same $\mu$ values. The fact that the system converges toward the expected steady-state solution is an important (and reassuring) result: under the turbulence prescriptions adopted here, the JED solution behaves as an attractor (in the dynamical sense). It is the outcome of a complex interplay between inward advection and outward turbulent diffusion of the large-scale magnetic field (see also \citealt{Sheikhnezami2012, Stepanovs2014}). The investigation of this process is postponed to future work.

These results, however, remain strongly dependent on the assumptions adopted to model MHD turbulence. Nevertheless, in full 3D global MHD simulations, where MRI-driven turbulence is captured,  inward advection of the large-scale magnetic field is also observed (see, e.g., \citealt{JacqueminIde2021}). However, a detailed comparison between the final steady-state outcome of 3D simulations of strongly magnetized (i.e., near-equipartition) accretion-ejection structures and analytical solutions is still lacking. Some attempts have been made in the context of GRMHD simulations of MADs (see, e.g., \citealt{Scepi2024b}), but these efforts suffer from the lack of detailed diagnostics on MHD turbulence (such as vertical profiles of anomalous transport coefficients) and on the jets launched by the disk (e.g., invariants, energetics, critical surfaces), and they have so far mostly been focused on radii very close to the black-hole (almost plunging region), where general relativistic effects start to come into play. A natural continuation of this work would be to extract turbulence properties from global 3D simulations in order to build physically motivated prescriptions and derive consistent scaling laws for implementation in 2.5D simulations and self-similar models.

While cold JEDs provide disk ejection efficiencies of $\xi \!\sim\! 0.01$, which is consistent with mildly relativistic jets observed on large scales in X-ray binaries (XrBs) or AGNs (see, e.g., \citealt{Petrucci2010}), such low values of $\xi$ imply jets that are too tenuous and too fast to account for those seen in young stellar objects (YSOs), which are denser and slower (see, e.g., \citealt{Garcia2001a,Garcia2001b,Ferreira2006b}). However, additional heating, either from UV and X-ray illumination by the central object or the innermost disk regions, or from local turbulent dissipation in the upper layers, is known to enhance $\xi$ by up to a factor of ten \citep{Casse2000b}. These magneto-thermal winds from JEDs lead to denser, slower, and still super-FM collimated outflows, which is consistent with various observational signatures, particularly in YSO jets \citep{Tabone2017,Tabone2020,Lee2021}, XrB winds \citep{Chakravorty2016,Chakravorty2023,Datta2024}, and post-AGB binaries \citep{Verhamme2024,DePrins2024}. It therefore becomes crucial to investigate thermodynamics in the context of strongly magnetized turbulent accretion disks, and to determine whether MRI-driven turbulence also contributes significantly to anomalous vertical heat transport (e.g., \citealt{Shakura1978,Scepi2024b}).

\section{Conclusion}

We performed axisymmetric numerical MHD simulations of accretion disks threaded by a large-scale vertical magnetic-field near equipartition. The disk is assumed to sustain MRI-driven turbulence, which is modeled in our 2.5D framework by introducing anomalous magnetic diffusivities and viscosity, and, for the first time, a turbulent magnetic pressure. All turbulent terms follow $\alpha$-type prescriptions that are consistent with current knowledge of MHD turbulence. Since the turbulent pressure has been shown to play a major role in accretion disks, this work represents a significant upgrade of $\alpha$ prescriptions for MHD turbulence in 2.5D simulations.

All simulations converged to the so-called JED solution, as described in \citet{Ferreira1997}, \citet{Zimniak2024} and references therein. A thorough comparison with self-similar models, both with and without turbulent magnetic pressure, shows that these models accurately reproduce the final stationary state of the simulations. This is the first time such a comparison has been successfully carried out, confirming that strongly magnetized (near equipartition) disks naturally lead to very-low-ejection efficiencies of $\xi \!\sim\! 0.01$ in the absence of additional heating. However, self-similarity is found to overestimate the hoop stress beyond the fast-magnetosonic point, causing the jet to recollimate toward the axis. It is unclear if such a recollimation occurs on much more distant scales \citep{Jannaud2023}, but both the analytical and numerical jets studied here follow a near-parabolic shape.

These are important results. On one hand, they confirm the validity of analytical self-similar solutions, which can serve as both benchmarks for 2.5D numerical simulations and as valuable tools for interpreting their outcomes. They also offer a robust framework for exploring new physical effects, such as heating in the upper disk layers or different turbulence prescriptions, and for easily exploring large parameter spaces.  On the other hand, 2.5D simulations will remain an essential approach for studying non-self-similar astrophysical systems, especially given the extremely large spatio-temporal scales that are still out of reach for 3D simulations. Nevertheless, a proper analysis of the MHD turbulence captured in fully converged global 3D simulations is urgently needed, in order to design physically motivated $\alpha$-type prescriptions for use in 2.5D models.

This work suggests that a JED behaves as an attractor, at least under the turbulence prescriptions adopted here (see also \citealt{Stepanovs2014}). If this result is dynamically confirmed, then hybrid disk configurations featuring an inner magnetically saturated JED region and an outer weakly magnetized WED region would likely arise naturally in systems with access to a substantial magnetic flux reservoir. Such a configuration, currently referred to as a JED-SAD structure in the literature, appears particularly promising for explaining transition disks in YSOs \citep{Combet2008,Combet2010,Martel2022} and the intriguing accretion–ejection hysteresis cycles observed in most XrBs \citep{Ferreira2006a, Marcel2018a, Marcel2018b, Marcel2022, Barnier2022}.

\begin{acknowledgements}
     NZ and JF acknowledge financial support from the ENIGMASS+ Labex and ATPEM program of French CNRS. CZ  acknowledges financial support from the Large Grant INAF-2024 ``Spectral Key features of Young stellar objects: Wind-Accretion LinKs Explored in the infraRed (SKYWALKER)''. All the computations presented in this paper were performed using the GRICAD infrastructure (\url{https://gricad.univ-grenoble-alpes.fr}), which is supported by Grenoble research communities.
\end{acknowledgements}

\bibliographystyle{aa}
\bibliography{bibliography}

\newpage

\appendix

\section{Initial and boundary conditions} \label{sect:A}

In the initial conditions we first set up a sligthly sub-Keplerian accretion disk threaded by a purely poloidal magnetic field. The initial disk structure is designed to reproduce a configuration as close as possible to a self-similar solution, without imposing a priori any specific accretion or ejection dynamics. The initial disk model is computed by looking for an equilibrium between gravity, rotation, Lorentz force and pressure gradients by solving the system of equations:
\begin{equation}
 \begin{aligned}
   \frac{\partial \left( P+P_{turb} \right)}{\partial R} & = -\frac{\rho GM}{R^2} + \frac{\rho v_\phi^2}{R} - J_\phi B_\theta \\
   \frac{1}{R}\frac{\partial \left( P+P_{turb} \right)}{\partial \theta} & =  \frac{\rho v_\phi^2 \cot\!\theta}{R} + J_\phi B_R \; ,
 \end{aligned}
 \label{eq:equilibrium}
\end{equation}
where the toroidal electric current $J_\phi$ is defined by the Maxwell–Ampère’s equation $\vec J = \nabla \times \vec{B} /\mu_0$. The system of equations (\ref{eq:equilibrium}) can be greatly simplified by looking for radial self-similar solutions. Assuming a vertically isothermal structure, the disk thermal pressure $P_d$ and density $\rho_d$ can be written as
\begin{equation}
 \begin{aligned}
 P_d & = P_{d0} \left(\frac{R}{R_0}\right)^{-5/2} h(\theta) \\
 \rho_d & = \rho_{d0} \left(\frac{R}{R_0}\right)^{-3/2} h(\theta) \sin\!\theta \; ,
 \end{aligned}
\end{equation}
where $P_{d0}$ and $\rho_{d0}$ are the midplane pressure an density at $R=R_0$ respectively. The disk constant thermal aspect ratio $\epsilon$ is defined as:
\begin{equation}
\epsilon = \left. \frac{c_s}{V_K}\right|_{z=0} = \left( \frac{P_{d0}}{\rho_{d0}} \frac{R_0}{GM} \right)^{1/2}  \; . 
\end{equation}
The initial poloidal disk magnetic field is set through the self-similar flux function as in \citet{Zanni2007}
\begin{equation}
\Psi_d = \frac{4}{3} B_{d0} R_0^2 \left(\frac{R}{R_0}\right)^{3/4} \frac{m^{5/4} \sin^2\!\theta}{\left[m^2\sin^2\!\theta +\cos^2\!\theta \right]^{5/8}} \; ,
\label{eq:psid}
\end{equation}
where $B_{d0}$ is the initial vertical midplane field at $R = R_0$ and $m$ is a dimensionless parameter that controls the field curvature above the disk equator. The components of the initial
poloidal field are defined as:
\begin{equation}
 \begin{aligned}
  B_R & = \frac{1}{R^2 \sin\!\theta} \frac{\partial \Psi_d}{\partial \theta} \\
  B_\theta & = - \frac{1}{R \sin\!\theta} \frac{\partial \Psi_d}{\partial R} \;. 
 \end{aligned}
\end{equation}
Notice that, since $\Psi_d \propto \sin^2\!\theta$ for $\theta \rightarrow \pi/2$, the field intensity has a singularity only for $R=0$ (a point that is not included in the computational domain), while it is vertical ($B_\theta = 0$) and with a finite value along the symmetry axis. The magnetic field $B_{d0}$ is customarily expressed as a function of the disk magnetization $\mu = B_{d0}^2/\mu_0 P_{d0}$ which is initially constant along the disk equatorial plane. Following the self-similar structure of the solution, the disk toroidal speed in the system of equations (\ref{eq:equilibrium}) is expressed as $v_\phi = \upsilon(\theta)\sqrt{GM/R}$. Given these assumptions the radial equilibrium in the system of equations (\ref{eq:equilibrium}) can be rewritten as
\begin{equation}
\begin{aligned}
\upsilon^2(\theta) = 1 & -\frac{5}{2}\epsilon^2\frac{1+\alpha_P\sqrt{\mu} f(\theta)}{\sin\!\theta}+ \\
& + \frac{\mu_0 J_\phi B_\theta R}{B_{d0}^2} \left(\frac{R}{R_0}\right)^{5/2} \frac{\mu \epsilon^2}{h(\theta)\sin\!\theta}
\end{aligned}
\label{eq:ups}
\end{equation}
providing an expression for the toroidal speed, while the transverse equilibrium becomes
\begin{equation}
\begin{aligned}
\left[1+\alpha_P\sqrt{\mu} f(\theta) \right] \frac{\partial h(\theta)}{\partial \theta}  = & \frac{\upsilon^2(\theta) h(\theta) \cos\!\theta}{\epsilon^2} - \\
& - \frac{\alpha_P\sqrt{\mu} f(\theta) \; h(\theta)\cos\!\theta}{\epsilon^2 x_t^2 \sin^3\!\theta} + \\
& + \frac{\mu_0 J_\phi B_R R}{B^2_{d0}}\left(\frac{R}{R_0}\right)^{5/2} \mu \; ,
\end{aligned}
\label{eq:ht}
\end{equation}
which is a first-order ordinary differential equation for the function $h(\theta)$ which can be integrated numerically with a boundary condition $h(\pi/2) = 1$ up to $h(\theta) \rightarrow 0$. Notice that the magnetic force terms in Eqs. (\ref{eq:ups}-\ref{eq:ht}) are a function of $\theta$ only and do not depend on $R$ due to the self-similar structure of the flux function Eq. (\ref{eq:psid}). Equation (\ref{eq:ups}) shows that the initial rotation of the disk at the midplane is sub-Keplerian with
\begin{equation}
\left. v_\phi \right|_{\theta = \pi/2} = \left[1-\frac{5}{2}\epsilon^2\left(1+\alpha_P\sqrt{\mu} \right) -\epsilon^2\mu \left(\frac{5}{4} + \frac{5}{3m^2} \right)\right]^{1/2} \sqrt{\frac{GM}{R}}
\label{eq:vphimid}
\end{equation}
and decreases towards the surface of the disk. Even if the accretion-ejection dynamics will develop as the simulations evolve, we initially set a radial velocity component $v_R$ defined by the induction equation so as to balance the diffusion of the magnetic field through the disk,
\begin{equation}
v_R = \frac{\eta_m J_\phi}{B_\theta}
\end{equation}
in order to provide an initial estimate of the accretion speed necessary to maintain the poloidal magnetic structure. The definition of the magnetic resistivity is provided in Section (\ref{sect:Ns}). This definition provides a value of the radial speed along the disk midplane
\begin{equation}
\left. v_R \right|_{\theta = \pi/2} = - \alpha_m \sqrt{\mu} \epsilon^2 \left(\frac{5}{4} + \frac{5}{3m^2} \right) \sqrt{\frac{GM}{R}} \; .
\end{equation}  
The initial toroidal magnetic field $B_\phi$ and the $v_\theta$ velocity component are set to zero. Inside the disk, the passive tracer $T_{\!r}$ is set to one, so that the material that will emerge from it to develop a disk-wind will have the same tracer value. 

On top of the disk we impose a hot atmosphere whose density and pressure radial profiles correspond to a spherically symmetric, isentropic (according the entropy definition given in Appendix A in \citet{Pantolmos2020}) transonic Parker-like thermal wind model. This solution is determined by the sound speed at the inner boundary, which we assume to be $c_s = 0.39 \sqrt{GM/R_0}$ and the density $\rho_{s0}$ at $R=R_0$. Initially we don't set the poloidal speed corresponding to this model, since it would determine the presence from the beginning of a supersonic flow that could strongly perturb the disk configuration. We let the stellar wind solution develop in time thanks to this initial profiles and suitable conditions on the inner boundary, so as to support an axial wind emerging from the inner boundary at latitudes larger than the accretion disk filling the disk-wind in the axial region. Therefore in the initial corona we assume all the velocity components and the toroidal field to be zero. We fill it with a potential, i.e. current-free ($J_\phi = 0$) and force-free, poloidal magnetic field determined by a self-similar flux function in the form
\begin{equation}
\Psi_s = \Psi_{s0} \left( \frac{R}{R_0}\right)^{3/4} \Phi(\theta) \; .
\end{equation}
The function $\Phi(\theta)$ is determined by numerically solving a second-order differential equation corresponding to the $J_\phi = 0$ requirement, imposing a regularity condition $ \Phi(\theta) \propto \sin^2\!\theta$ for $\theta \rightarrow \pi/2$, so that the magnetic field is vertical but not singular along the symmetry axis. This is the same procedure adopted in \citet{Jannaud2023}. In the initial atmosphere the tracer $T_{\!r}$ is set to zero. 
The boundary between the initial disk and the hot ``stellar'' atmosphere is set at the inner radius $R_0$ at a latitude where the thermal pressures of the two media are equal. Since the atmosphere pressure does not follow the same self-similar radial profile of the disk, this angle varies slightly with radius (by about one to two grid cells), but these variations are neglected, and the angle is assumed to be constant. The requirement that the disk and the coronal magnetic fields have the same value and the same inclination at this constant latitude is used to fix both the coronal field intensity $\Psi_{s0}$ and the initial inclination of the disk field controlled by the parameter $m$, which for our simulations assumes typical values in the range $m \approx 0.33 - 0.39$. 

On the rotation axis and the disk midplane we assume customary axial-symmetric and equatorial-symmetric boundary conditions. On the inner boundary $R=R_0$ the polodial magnetic field is set by extrapolating the $B_\theta$ component into the ghost cells using the initial power-law profile $\propto R^{-5/4}$. The radial component $B_R$ is determined by the $\nabla\cdot\vec{B} = 0$ condition. This boundary condition allows the possibility of advecting and/or diffusing the poloidal magnetic flux in and out of the inner boundary consistently with the development of the accretion-ejection dynamics. For the other quantities we consider two types of boundary conditions, one for the outflowing material (the accretion disk) and another for the inflowing one (the axial wind). For the accretion disk we extrapolate all the quantities into the ghost zones using the same self-similar power-laws used to compute the initial conditions, namely $\rho \propto R^{-3/2}$, $P\propto R^{-5/2}$, all the velocity components $\propto R^{-1/2}$ and the toroidal field $B_\phi \propto R^{-5/4}$. For the thermally-driven wind emerging from the inner boundary we fix the pressure and density to the profiles that we used to initialize the atmosphere, so as to guarantee the proper pressure gradient to support a subsonic injection. The poloidal speed is set to be parallel to the magnetic field and it is copied from the first cell inside the domain, with an extra check to ensure that the injection speed stays subsonic. The toroidal magnetic field component is linearly extrapolated into the ghost cells, using a three-point stencil inside the domain to determine the linear slope. To determine the toroidal speed we also require that the magnetic surfaces of the inner axial wind rotate with an angular speed $\Omega_\star$
\begin{equation}
r\Omega_\star = v_\phi - v_p \frac{B_\phi}{B_p} \; ,
\end{equation}
where $v_p$ and $B_p$ are the poloidal speed and magnetic field, and $\Omega_\star$ is set to be equal to the initial value of the angular speed of the accretion disk at its inner orbit on the equatorial plane, see Eq. (\ref{eq:vphimid}). The value of the passive tracer in the first cell of the computational domain, which is equal to one for the accretion disk and zero for the axial wind, is used to decide which boundary condition must be applied.
The outer radial boundary should allow the plasma to leave the computational domain trying to minimize any feedback. On this boundary all the quantities are extrapolated outwards using power laws whose exponents are calculated on a three-point stencil inside the computational domain. 

\section{Temporal convergence}\label{sect:B}

There are three ways to define a steady-state regime in accretion–ejection theory. The first approach is to define a global steady-state domain $\mathcal{D}_{ss}$ containing all magnetic surfaces that cross the FM critical surface, the latter marking the limit beyond which the jet can no longer exert a causal influence on the disk.

\noindent Figure~\ref{fig:fig2} shows the time evolution of the anchoring radius $R_{ss}$ of the outermost magnetic field line crossing the FM critical surface. At $t \!=\! 1000$, global stationarity is reached up to a radius $R_{ss} \!=\! [13.3, 4.5, 2.3]$ for $\alpha_P \!=\! [0,1,2]$.

A second, less stringent (though potentially less reliable) definition of stationarity can also be adopted. The previous definition relies on a demanding global criterion, requiring a complete reorganization of the system in both radial and vertical directions. In practice, a locally steady-state regime $\mathcal{D}_{ss,\ell}$ can be established at larger radii once transient phenomena have dissipated. From this perspective, a locally steady region can be defined as the domain where the accretion rate $\dot{M}_{acc} \!=\! 4 \pi R^2 \int_0^{\theta_d} \rho v_R sin(\theta) d \theta$ becomes time-independent.

\begin{figure}
    \centering
    \includegraphics[width=1.0\columnwidth]{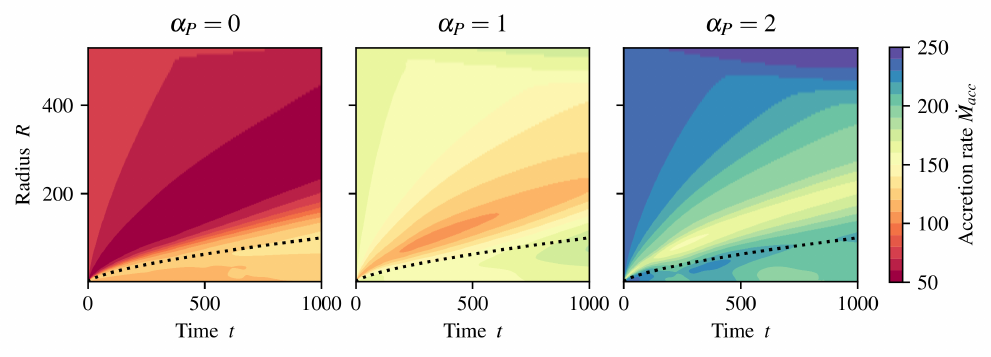}
    \caption{Space-time diagram of the disk accretion $\dot M_{acc}(R)$ for each simulation. The dashed curve follows the keplerian $r^{3/2}$ scaling.}
    \label{fig:figB1}
\end{figure}

\noindent The temporal and radial evolution of $\dot{M}_{acc}$ for each simulation is shown in Fig.~\ref{fig:figB1}. A region gradually forms in which the accretion rate tends toward a quasi-stationary behaviour, although residual fluctuations remain. These fluctuations are linked to ongoing dynamical adjustments or to perturbations that are only partially dissipated. The extent of this locally steady region follows a $t^{2/3}$ power law, consistent with the characteristic accretion time in the disk $t_{acc} \!=\! r/v_r \propto \Omega_K^{-1} \propto r^{\frac{3}{2}}$.

\noindent At $t \!=\! 1000$, local stationarity is reached up to a radius $R_{ss,\ell} \!=\! 100$, regardless of $\alpha_P$. This value exceeds $R_{ss}$ by at least an order of magnitude, potentially allowing, in principle, the analysis to be extended over a much larger region.

However, in an accretion–ejection context, the relevant steady-state region is the one where the MHD invariants $(\kappa, \omega_*, \lambda, e)$, measured along magnetic surfaces, remain constant in time. This region is, however, not computed at each time $t$ in this study, as the numerical cost of the complex interpolations it requires would be prohibitive: at each time step, one would need to trace the coordinates of several hundred magnetic field lines in the domain and interpolate all invariants along them to determine the extent of the steady region.

\begin{figure}
    \centering
    \includegraphics[width=0.9\columnwidth]{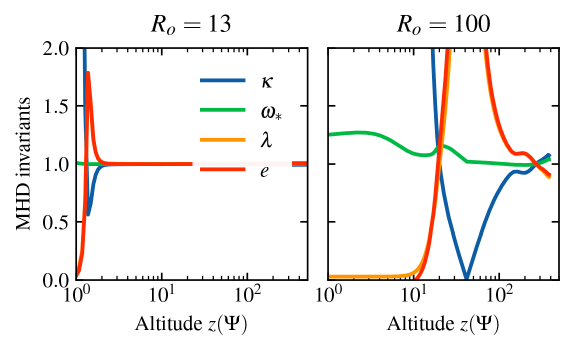}
    \caption{Profiles of the four MHD invariants $\kappa, \omega_*, \lambda, e$ along a magnetic surface anchored at $r_o \!=\! R_{ss}= 13$ (left) and $r_o \!=\! R_{ss,l} \!=\! 100$ for the $\alpha_P \!=\! 0$ simulation at $t \!=\! 1000$.}
    \label{fig:figB2}
\end{figure}

\noindent Nevertheless, we can compute these invariants only at $t \!=\! 1000$, along magnetic field lines anchored to $R_{ss}$ and $R_{ss,\ell}$, to assess the validity of the global and local steady-state domains. These invariants, normalized to their values at the jet base and measured for the $\alpha_P \!=\! 0$ simulation, are shown in Fig.~\ref{fig:figB2}. The results indicate that the invariants along the field line anchored at $R_{ss} \!=\! 13$ remain constant, confirming the presence of a steady-state regime in $\mathcal{D}_{ss}$. In contrast, those measured along the line anchored at $R_{ss,\ell} \!=\! 100$ show significant variations, incompatible with any form of stationarity. We therefore conclude that $\mathcal{D}_{ss,\ell}$ does not provide a sufficiently reliable basis for a steady-state analysis of accretion–ejection properties. Only $\mathcal{D}_{ss}$ meets this criterion, and it is this domain that we adopt throughout the paper.

\section{Supplementary tests} \label{sect:C}

\subsection{Resolution study}

To quantify the impact of numerical diffusion, we carried out a series of seven additional simulations in which the spatial resolution was deliberately degraded. The input physical parameters were kept fixed, $\varepsilon \!=\! 0.1$, $\alpha_m \!=\! \mathcal{P}_m \!=\! \chi_m \!=\! 1$, $\alpha_P \!=\! 0$ and $\mu_i \!=\! 0.6$, while the latitudinal resolution was adjusted to span between 2 and 16 grid cells across the disk thickness. Two combinations of Riemann solver and slope limiter were considered: HLLD with van Leer, which is less diffusive, and HLL with Minmod, which is more diffusive. For each simulation, we measure the magnetic lever arm $\lambda$ within the steady-state domain, as an indicator of the mass loss from the disk, since it is more straightforward to compute than $\xi$.

\begin{figure}
    \centering
    \includegraphics[width=0.7\columnwidth]{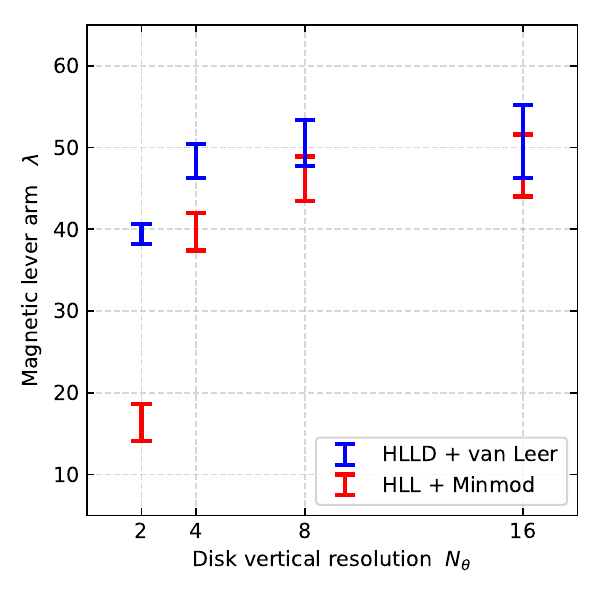}
    \caption{Magnetic lever arm $\lambda$ as a function of vertical resolution in the accretion disk. The blue points correspond to simulations with low numerical diffusion (HLLD solver and van Leer slope limiter). The red points correspond to simulations with high numerical diffusion (HLL solver and Minmod slope limiter). For each vertical resolution, the interval in $\lambda$ indicates the range of values measured across the steady-state domain.}
    \label{fig:figC1}
\end{figure}

The results, measured at $t \!=\! 1000$, are shown in Fig.~\ref{fig:figC1}. For each vertical resolution, the interval in $\lambda$ indicates the range of values measured across the steady-state domain. We find that $\lambda$ increases with vertical resolution: in other words, the mass loading of the jet decreases as resolution improves, confirming the presence of significant numerical diffusion when the resolution is too low. This effect can reduce $\lambda$ by up to a factor of three. Furthermore, while the difference between the two numerical schemes remains modest at high resolution, it becomes substantial once the vertical resolution is reduced. It is therefore crucial, whenever possible, to adopt sufficient resolution in order to limit such non-physical effects.

\begin{figure}
    \centering
    \includegraphics[width=0.9\columnwidth]{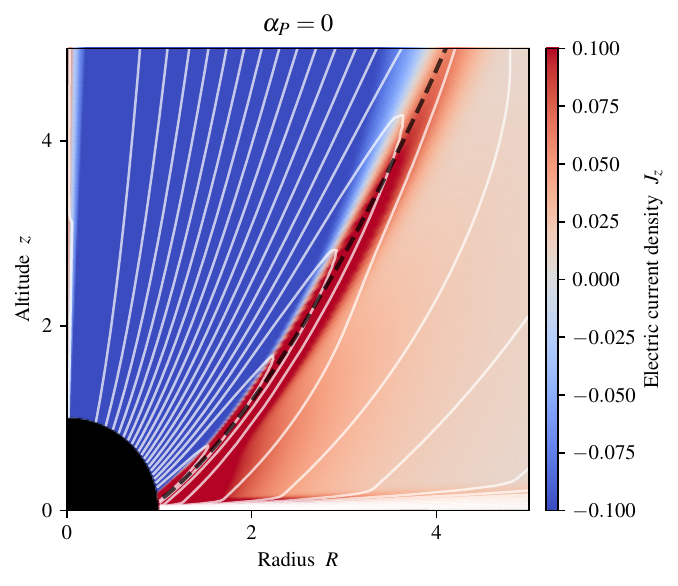}
    \caption{2D visualization at $t \!=\! 1000$ for the $\alpha_P \!=\! 0$ simulation, zoomed on the inner region. White lines show isocontours of the poloidal current $I \!=\! 2 \pi r B_{\phi} / \mu_0$, while the dashed black line marks the spine boundary. The background colormap represents the vertical component of the current density $J_z$ in arbitrary units.}
    \label{fig:figC1bis}
\end{figure}

Thus, the differences in ejection rates $\xi$ between the present work and previous studies are most likely caused by numerical diffusion associated with low resolution. Another important factor, however, is the topology of the electric current. Fig.~\ref{fig:figC1bis} presents a 2D zoom of the inner regions, showing isocontours of the poloidal current $I \!=\! 2 \pi r B_{\phi} / \mu_0$ at $t \!=\! 1000$ for the $\alpha_P \!=\! 0$ simulation. The vertical component of the current density $J_z$, is displayed in color, while the boundary between the spine and disk-driven material is marked in black.

The current flows along the axis (return current), crosses the disk through its inner edge, and emerges again through its surface, thereby closing the electric circuit and forming a characteristic "butterfly"-like structure. This configuration is typical of magneto-centrifugal acceleration \citep{Ferreira1997, Jannaud2023}. The main difference between studies lies in the path of the return current: in our case, it passes through the inner edge of the disk, whereas in other works it sometimes penetrates the disk directly through its surface (e.g. \citealt{Zanni2007}), which can significantly increase the jet mass loading. This topology is, of course, partly determined by boundary conditions. In our simulations, the transition between positive and negative $J_z$ nearly coincides with the spine boundary, suggesting that the presence of the return current along the axis may be partly induced by the spine itself. However, it remains unclear whether resolution also affects the return current.

\begin{figure*}[t]
    \centering
    \includegraphics[width=1.0\textwidth]{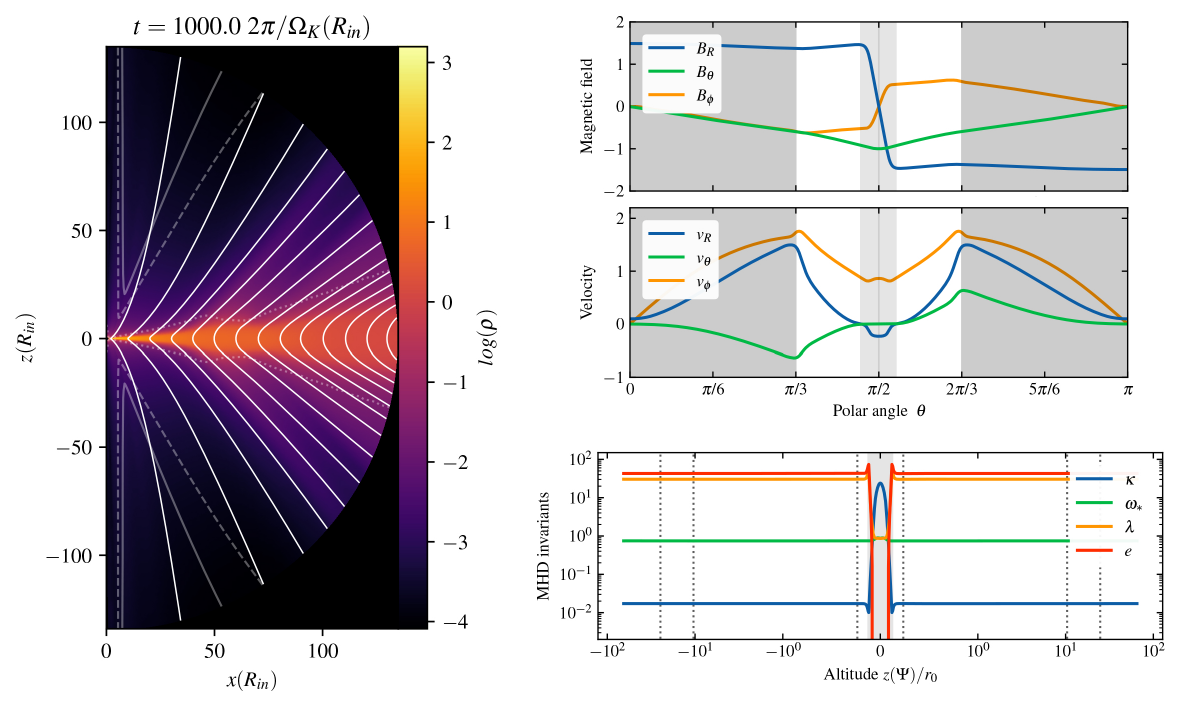}
    \caption{Left: 2D visualization at $t \!=\! 1000$ for the simulation performed over the interval $[0, \pi]$. Dashed lines, with increasing opacity, represent the slow magnetosonic (SM), Alfvén, and fast magnetosonic (FM) critical surfaces, respectively. White lines indicate magnetic field lines anchored at $R \!=\! 2$, $10$, $20$, $30$, up to $130$. The background colormap shows the density in arbitrary units.
    Upper right: Latitudinal profiles of the magnetic field components $B_R$, $B_{\theta}$, $B_{\phi}$ (normalized to $B_z$ at the disk midplane), and components of the velocity $v_R$, $v_{\theta}$, $v_{\phi}$ (normalized to the Keplerian velocity $v_{K}$ at the disk midplane), measured at $R \!=\! 2.0$ and $t \!=\! 1000$. The central gray-shaded region corresponds to the disk, the white region to the disk wind, and the outer gray-shaded regions to the spine.
    Lower right: Profiles of the four MHD invariants $\kappa, \omega_*, \lambda, e$ along the magnetic field line anchored at $R \!=\! 2.0$, measured at $t \!=\! 1000$. The shaded area indicates the resistive MHD region, while the vertical dashed lines mark the positions of the SM, A, and FM critical surfaces.}
    \label{fig:figC3}
\end{figure*}

\subsection{JED simulation on $[0,\pi]$ domain}

Are the results still valid for simulations spanning the full polar range $\theta \in [0,\pi]$? We addressed this by performing a simulation over the same grid spacing as before but restricted in radius (for computational efficiency), with $R \in [1,135]$ and $\theta \in [0,\pi]$. The initial parameters are identical to those used previously, namely $\mu_i \!=\! 0.6$, $\varepsilon \!=\! 0.1$, $\alpha_m \!=\! \mathcal{P}_m \!=\! \chi_m \!=\! 1$, with no turbulent magnetic pressure ($\alpha_P \!=\! 0$).

Figure~\ref{fig:figC3} shows snapshots at $t \!=\! 1000$. The two sides of the midplane are strictly identical in density, magnetic field, velocity, critical surfaces and MHD invariants. This perfect north–south symmetry stands in contrast to other studies where the disk midplane is not a symmetry plane, for instance, the axisymmetric 2D simulations of \citet{Tu2025}, which report jets that deviate from perfect symmetry. In their model, jets are launched by avalanche accretion streams that undergo frequent magnetic reconnection. These events are highly time-variable and can occur first on one side of the disk, changing the local mass loading and magnetic field geometry there before anything similar happens on the other side. This acts like an internal "random" forcing: it is not prescribed in the equations, but it emerges naturally from the dynamics. All of this variability comes from the resolved MHD processes and the reconnection captured at their resolution. The result is that, at any given moment, the two jets are not exactly identical, even though their average properties remain similar. In 3D, additional instabilities and turbulence would provide more such random perturbations, making perfect symmetry even harder to maintain.

\end{document}